\documentclass[reprint, prx,  superscriptaddress]{revtex4-2}

\usepackage{verbatim}
\usepackage{graphicx}
\usepackage{color}
\usepackage{amsmath}
\usepackage{amssymb}
\usepackage{braket}
\usepackage{enumerate}
\usepackage{multirow}
\usepackage{colortbl}
\definecolor{Gray}{gray}{0.3}
\usepackage{wrapfig}
\usepackage[latin1]{inputenc}
\usepackage[dvipsnames]{xcolor}
\usepackage{soul}
 
\DeclareMathAlphabet{\mathcal}{OMS}{cmsy}{m}{n}
\newcommand{\be}{\begin{equation}}
\newcommand{\ee}{\end{equation}}

\newcommand{\mean}[1]{\langle #1 \rangle}

\newcommand{\fig}[1]{Fig.~\ref{#1}}

\newcommand{\Fig}[1]{Figure~\ref{#1}}

\newcommand{\eq}[1]{Eq.~\eqref{#1}}
\newcommand{\eqs}[2]{Eqs.~\eqref{#1} and \eqref{#2}}

\newcommand{\stn}[1]{Sec.~\ref{#1}}

\newcommand{\bea}{\begin{eqnarray}}
\newcommand{\eea}{\end{eqnarray}}
 
\newcommand{\ba}{\begin{array}}
\newcommand{\ea}{\end{array}}

\newcommand{\bl}{\begin{flalign}}
\newcommand{\enl}{\end{flalign}}

\newcommand{\op}[2]{| {#1} \rangle \langle {#2}|}

\newcommand{\mycomment}[1]{}

\usepackage{makecell}

\usepackage[defaultcolor=magenta]{changes} 
\setaddedmarkup{\color{blue}#1} 

\begin{document}

\title{Decoherence dynamics in molecular qubits: Exponential, Gaussian and beyond}

\author{Ignacio Gustin}
\affiliation{
    Department of Chemistry, University of Rochester, Rochester, New York 14627, USA
    }
  \author{Xinxian Chen}
\affiliation{
    Department of Chemistry, University of Rochester, Rochester, New York 14627, USA
   }

\author{Ignacio Franco}
\email{ignacio.franco@rochester.edu}
\affiliation{
    Department of Chemistry, University of Rochester, Rochester, New York 14627, USA
    }
\affiliation{
    Department of Physics, University of Rochester, Rochester, New York 14627, USA
    }

\date{\today}
\begin{abstract}
 In this work, we examine how the structure of system-bath interactions can determine commonly encountered temporal decoherence patterns, such as Gaussian and exponential decay, in molecular and other qubits coupled to a thermal bosonic bath. The analysis, based on a pure dephasing picture that admits analytical treatment, shows that decoherence, in general, is neither purely Gaussian nor exponential but rather the exponential of oscillatory functions, with periods determined by the bath's frequencies.  For initially unentangled qubit-bath states  Gaussian decay is always present at early times. It becomes increasingly dominant with increasing temperature, qubit-bath interaction strength, and bath correlation time. Initial system-bath entanglement that arises due to displacement in the position of the bath states preserves the Gaussian decay. By contrast, strict exponential decay arises only in very specific models that we isolate. Still, it becomes dominant for times longer than the bath correlation time or for early times when there is initial entanglement due to momentum displacement of the bath states. For molecular electronic decoherence, the long-time exponential regime plays a limited role as it emerges after most coherence is lost. Thus, the Gaussian decay provides a more suitable (albeit imperfect) model of such decoherence. Further, we discuss the connection between electronic decoherence dynamics and electronic spectroscopic lineshape theory, where Gaussian spectral peaks correspond to Gaussian coherence decay and Lorentzian peaks correspond to exponential coherence decay. We find that Gaussian spectral peaks, usually associated with inhomogeneous broadening, can emerge from the entangling unitary system-bath  dynamics even when there is no inhomogeneity in the initial conditions.

\end{abstract}
\keywords{Quantum Dynamics, Open Quantum Systems, Molecular qubits}

\maketitle

\section{Introduction}
Quantum coherence refers to the ability of matter to sustain superposition of states as required for matter to exhibit quantum features such as its ability to interfere or be entangled. For this reason, quantum coherence is an essential element in quantum information science (QIS) and necessary for exerting quantum control over matter.\cite{Wasielewski_2020,McArdle2020,Acin_2018,Cappellaro_2017,Brumer2012,ricebook} Molecules, in particular, are highly compact and configurable quantum systems. They offer a range of chemically tunable energy levels across the UV/Vis (electronic/vibronic), infrared (vibrational), and microwave (rotational/spin) regions of the electromagnetic spectrum, enabling quantum operations from femtoseconds to milliseconds. 

Despite this promise, molecular qubits are currently not primary candidates for QIS. This is because molecular quantum coherence is particularly fragile to decoherence (or quantum noise) processes that arise due to the unavoidable and uncontrollable interactions of the molecules with their surrounding environment or bath. \cite{freedman2022,Schlosshauer_2007,Breuer2002,Caram2022,viola1999,Gustin2023}  In fact, electronic ($\sim$ 10 fs) and vibrational ($\sim$ 1000 fs) decoherence in molecules is typically remarkably fast.\cite{Gustin2023,Hwang2004,Fleming_1990} 

To harness the potential of molecular qubits in QIS, it is important to identify robust molecular design principles to generate quantum subspaces with protected quantum coherence.\cite{Freedman_2015,Freedman_2021,Caram2022,freedman2022,wang2019turning,dickerson2021} Achieving this requires understanding how the molecule-bath interactions dictate the qubit decoherence dynamics, as needed to inform strategies to tune the bath to control decoherence. Understanding decoherence is also central in our elementary description of photophysics\cite{Mukamel_book}, photochemistry\cite{Engel_2019}, multidimensional optical spectroscopies\cite{Harel_2019}, in designing quantum control strategies\cite{Brumer2012,ricebook}, and in developing methods to approximately capture quantum molecular dynamics.

Throughout, we focus on pure dephasing processes that arise due to the interaction of a qubit with a thermal harmonic bath for which the decoherence function can be isolated analytically.\cite{unruh1995,Breuer2002,palma1996} While the approach does not capture relaxation, it remains highly informative of decoherence in molecular qubits where the pure dephasing processes usually occur on faster time scales than overall relaxation, thereby dominating the decoherence dynamics.\cite{brinks2014ultrafast, von1997laser}
In turn, the harmonic description of the bath is widely applicable because any system-bath problem can be rigorously mapped onto a system linearly coupled to a harmonic oscillator environment, provided the interaction can be dealt with up to second order in perturbation theory.\cite{Feynman1963, Caldeira1983, Leggett1987, Caldeira1993, wiethorn2023beyond} This situation is expected in molecular qubits where qubit-bath interactions are usually diluted over a macroscopic number of degrees of freedom.\cite{lunghi2019phonons,garlatti2023critical,lunghi2022toward} 
While our focus is on molecular systems, the pure-dephasing model in harmonic baths is extensively used to model decoherence in other qubit platforms\cite{Vezvaee2024,paz2017multiqubit,kwiatkowski2020influence,cywinski2008enhance}, making our insights of broad applicability.

A central quantity in this analysis is the spectral density, $J(\omega)$, characterizing the nuclear bath's  frequencies, $\omega$, and their coupling strength with the qubit.  Here, we show how the structure of the spectral density leads to commonly used and observed temporal decoherence decay patterns in molecular qubits, such as Gaussian and exponential decay. We find that decoherence, in general, is neither strictly Gaussian nor exponential but the exponential of oscillatory functions with periods determined by the bath's  frequencies. For initially unentangled qubit-bath states or initial entanglement due to displacement in the position of the bath states, we find that Gaussian decay is always present at early times and can become dominant as the temperature and the molecule-bath interactions strength. For initially unentangled states, the Gaussian regime also becomes increasingly dominant as the correlation time of the bath increases. In turn, we find that strict exponential coherence decay only occurs for very specific shapes of the spectral density that we isolate. However, it can dominate for times longer than the bath correlation time or for early times when there is initial entanglement due to displacement in the momentum of the bath states. While other models of initial qubit-bath entanglement can lead to different temporal decoherence patterns, the models employed here clearly demonstrate that initial qubit-bath entanglement, while often neglected in decoherence studies, can significantly affect the decoherence dynamics.  

We then investigate the relevance of Gaussian and exponential decoherence decay models in realistic systems, such as electronic decoherence in thymine derivatives in water at 300 K, where the spectral densities for initial separable system-bath states are known.\cite{Gustin2023} We observe that the Gaussian decay is dominant at early times. However, it overestimates the overall decoherence by a factor of $\sim$ 2. By contrast, the exponential decay is only observed after most of the molecular coherence has been lost. Thus, the Gaussian decoherence model provides a more suitable (albeit imperfect) description of electronic decoherence dynamics for molecules in condensed-phase baths.

Last, we reexamine the connection between decoherence and the theory of spectroscopic lineshapes. It is well known that electronic decoherence patterns and time scales can be estimated from absorption and emission lineshapes in the pure dephasing limit.\cite{schatz2002quantum,skinner1986pure,tokmakotime} Specifically, Lorentzian-shaped spectral peaks, referred to as the homogeneous limit, indicate exponential coherence decay. By contrast, Gaussian-shaped spectral peaks, known as the inhomogeneous limit, indicate Gaussian coherence decay. It is commonly believed that the homogeneous component captures dynamic processes intrinsic to the molecular system, while the inhomogeneous component emerges only from ensemble averaging effects. As a result, only the homogeneous part is typically considered to represent actual system-bath entanglement processes. Here, we demonstrate that Gaussian spectral peaks, generally associated with inhomogeneous effects, can arise from the entangling unitary system-bath dynamics even when there is no inhomogeneity in the initial conditions.

The paper is structured as follows. In Sec. \ref{sec:decoherence-functions}, we present the theoretical background for quantum decoherence functions. Secs. \ref{sec:Results} A-E, are focused on initially unentangled qubit-bath states. Specifically, in Secs. \ref{sec:Results} A-C, we discuss the requirements on the spectral density that lead to oscillatory, strictly exponential, and Gaussian temporal decay. In Sec. \ref{sec:Results} D, we examine the connection between decoherence and spectroscopic lineshapes.  In Sec. \ref{sec:Results} E, we investigate electronic decoherence in realistic molecules. Finally, in Sec. \ref{sec:Results} F, we examine the role of initial entanglement due to displacement in momentum and position of the bath states in the decoherence dynamics. We summarize our main findings in Sec. \ref{conclusions}.

Overall, this manuscript provides an analysis of coherence behavior crucial in quantum information problems that draws inspiration from the rich body of work on quantum dynamics and spectroscopy in physical chemistry. We recapitulate known results such as early-time Gaussian and long-time exponential coherence decay and isolate interesting and useful conditions for exclusive Gaussian or exponential decay and their applicability in electronic decoherence, the influence of initial-state entanglement in coherence loss, and demonstrate that Gaussian lineshapes in spectroscopy can arise from unitary entangling system-bath dynamics beyond the inhomogeneous limit. 

\section{Theoretical Background: Quantum decoherence function}\label{sec:decoherence-functions}
We divide the molecular Hamiltonian $H=H_{\text{S}}+H_{\text{B}}+H_{\text{SB}}$ into a system (qubit) $H_{\text{S}}$, a bath $H_{\text{B}}$, and their interaction $H_{\text{SB}}$. The pure dephasing condition requires that $[H_{\text{S}},H_{\text{SB}}]=0$, thus guaranteeing that the system energy is conserved even in the presence of a bath. While this condition is not strictly satisfied in molecules, the pure dephasing effects still dominate when there is a disparity in time scales between the dephasing and subsequent relaxation. That is the common case when energy dissipation occurs at a much slower rate than pure dephasing phenomena.\cite{Wen2018,Nitzan2006} For this reason, the pure dephasing limit has been useful in understanding electronic decoherence in molecules \cite{Gustin2023,Gu_2019}, lineshapes in laser spectroscopy \cite{kubo1969,Mukamel_book}, vibrational dephasing in solvents\cite{Joutsuka2016} and the central spin problem.\cite{yang2016quantum}

The system-bath dynamics can be described by the Liouville-von Neumann (LvN) equation
\begin{equation}
     i\frac{d}{dt}\tilde{\rho}(t)=\left [\tilde{H}_{\text{SB}}(t),\tilde{\rho}(t)\right],
\end{equation}
where $\tilde{O}(t) = U_{0}^{\dagger}(t)\hat{O} U_{0}(t)$ is the operator $\hat{O}$ in the interaction picture of $H_{0}=H_{\text{S}}+H_{\text{B}}$, and $U_{0}(t)=\exp (-i H_{0}t)$. We use atomic units throughout, where $\hbar=1$. As customary in open quantum dynamics, we assume that the system and bath are not correlated at initial time (we relax this assumption in Sec. \ref{sec:entangled}). The initial total density matrix can thus be written as 
\begin{equation} \label{eq:initial-rho}
    \rho(0) = \rho_{\text{S}}(0)\otimes\rho_{\text{B}}(0),
\end{equation}
where $\rho_{\text{S}}$ is the reduced density matrix of the system and $\rho_{\text{B}}$ that of the bath. The formal solution to the LvN equation is:
\begin{equation} \label{eq::Density_matrix}
    \tilde{\rho}(t)=\tilde{U}(t)\rho(0)\tilde{U}^{\dagger}(t),
\end{equation}
where $\tilde{U}(t)= \mathcal{T} \exp \left \{  -i \int_{0}^{t} d\tau \tilde{H}_{\text{SB}}(\tau) \right \}$ is the propagator in the interaction picture and $\mathcal{T}$ is the time-ordering operator. 

For pure dephasing, the system-bath interaction term in the Hamiltonian 
 can be written as $H_{\text{SB}}=\sum_{j}\op{j}{j}\otimes B_{j}$, where $\{\ket{j}\}$ are the  eigenstates of $H_{\text{S}}$, and $B_{j}$ is a bath operator. It follows that
 $\tilde{H}_{\text{SB}}(t)=\sum_{j}\op{j}{j}\otimes \tilde{B}_{j}(t)$  and
\begin{equation} \label{eq::Evolution-Operator}
    \begin{aligned}
    \tilde{U}(t)&= \mathcal{T} \sum_{n=0}^{\infty} \frac{(-i)^{n}}{n!} \left( \int_{0}^{t}d\tau\sum_{j}\op{j}{j}\otimes \tilde{B}_{j}(\tau)\right)^{n}\\
    &= \sum_{j}\op{j}{j}\otimes \mathcal{T} \sum_{n=0}^{\infty} \frac{(-i)^{n}}{n!} \left( \int_{0}^{t}d\tau \tilde{B}_{j}(\tau)\right)^{n}\\
    &= \sum_{j}\op{j}{j}\otimes V_{j}(t),
    \end{aligned}
\end{equation}
where we have defined $V_{j}(t)=\mathcal{T}\exp \left \{  -i \int_{0}^{t} d\tau \tilde{B}_{j}(\tau)\right\}$.  Inserting \eqs{eq:initial-rho}{eq::Evolution-Operator} into \eq{eq::Density_matrix}, and tracing out the bath degrees of freedom ($\text{Tr}_{\text{B}}[\cdots]$) leads to 
\begin{equation}
    [\tilde{\rho}_{\text{S}}]_{j i}(t)=\bra{j}\text{Tr}_{\text{B}}[\tilde{\rho}(t)]\ket{i}= [\tilde{\rho}_{\text{S}}]_{ji}(0)\Phi_{ji}(t)
\end{equation}
for the off-diagonal part of the reduced density matrix of the system. Here,
\begin{equation}\label{eq::Decoherence-Function}
    \Phi_{ji}(t)=\text{Tr}_{\text{B}}\left[\rho_{\text{B}}(0) V_{i}^{\dagger}(t) V_{j}(t)\right]=\left \langle V_{i}^{\dagger}(t) V_{j}(t)\right \rangle
\end{equation}
is the quantum decoherence function characterizing decoherence between states $\ket{j}$ and $\ket{i}$ due to system-bath interactions.

The decoherence function can be expressed as $\Phi_{ji}(t)=\exp{\left \{ 
-\chi(t)+i\phi(t) \right\}}$,\cite{paz2017multiqubit,kwiatkowski2020influence,mukamel1985fluorescence,Mukamel_book} where $\chi(t)$ and $\phi(t)$ are real functions. In this paper, we focus on the magnitude of the decoherence function $|\Phi_{ji}(t)|=\exp{\left \{ -\chi(t)\right \}}$ as it signals coherence loss since $| [\tilde{\rho}_{\text{S}}]_{j i}(t)|=| [\tilde{\rho}_{\text{S}}]_{ji}(0)||\Phi_{ji}(t)|$. For Gaussian environments this does not lead to information loss as $\chi(t)$ and $\phi(t)$ are related through the fluctuation-dissipation theorem, making $\chi(t)$ sufficient to extract all relevant information.\cite{Vezvaee2024}

\section{RESULTS AND DISCUSSIONS} \label{sec:Results}

Gaussian and exponential temporal coherence decay have been identified as common models for decoherence. For example, exponential coherence decay is usually used for spin-1/2 chains in Markovian baths\cite{cai2013}, or more generally, in the long-time limit\cite{knight1976,burgarth2017,unruh1989,paz1993,zurek2003,xu2019} with discussions in the contexts of Gaussian stochastic models\cite{tokmakotime}, spectroscopy\cite{Mukamel_book, skinner1986pure,hsu1984thermal}, the theory of liquids\cite{hansen2013theory}, and in quantum information science.\cite{palma1996,unruh1995} However, it is unclear if this long-time limit is of relevance in molecular-based qubits. Does the exponential regime emerge when there is still appreciable coherence in the molecular system? Moreover, the specific conditions that lead to strict exponential coherence decay have not been clarified. By contrast, the Gaussian form dominates at the early stage of decoherence\cite{Gu_2019,Bing2018,Wen2022}, which can be seen as arising due to the quantum Zeno effect.\cite{von2018,misra1977,facchi2008} This initial Gaussian decay is well-known, appearing in theories of early-time decoherence time scales\cite{Bing2017, Bing2018} and the theory of line-broadening functions in spectroscopy.\cite{tokmakotime, Mukamel_book, hsu1984thermal, kubo1969, kubo1962} However, it remains unclear the regime of validity of this Gaussian decay in molecular-based qubits and as a function of qubit-bath interaction strength, temperature, and bath's correlation time. 

More recently, studies based on perturbation theory and semiclassical analyses\cite{yan2022} have proposed that the decay of coherence is a convolution of Gaussian and exponential decay.
However, for molecular-based qubits where a molecular transition is strongly coupled to a few selected vibrational modes and weakly coupled to a macroscopic number of solvent/lattice modes, the correct temporal pattern of the decoherence remains unclear.

In this section, we seek to clarify how these Gaussian and exponential temporal decay patterns arise as a function of the structure of the bath or, more precisely, its spectral density and their relevance in electronic decoherence in molecules.

\subsection{Decoherence, in general, is neither Gaussian nor exponential}\label{sec:first}

The displaced harmonic oscillator is the standard model used to understand molecular qubit decoherence.\cite{Page1981Separation,shreve1995Thermal,Tannor1982Poly,Myers1982Excited} In the pure dephasing limit, the model is defined by the Hamiltonian  
\begin{equation} \label{eq:Molecular-Hamiltonian}
    H=H_{0}\op{0}{0}+H_{1}\op{1}{1},
\end{equation}
where $\ket{0}$ and $\ket{1}$ denote the ground and excited qubit state. Here, $H_{\text{0}}=\sum_{\alpha}\left( \frac{p_{\alpha}^2}{2m_{\alpha}}+\frac{1}{2}m_{\alpha}\omega_{\alpha}^{2}x_{\alpha}^{2} \right)$ is the ground-state bath Hamiltonian, where $x_{\alpha}$ and $p_{\alpha}$  are the position and momentum operators of the $\alpha$-th bath mode with mass-weighted frequency  $\omega_{\alpha}$. In turn, the excited-state Hamiltonian, $H_{1}$, consists of the same set of bath modes but displaced in conformational space, i.e., $H_{\text{1}}=\omega_{01}+\sum_{\alpha}\left( \frac{p_{\alpha}^2}{2m_{\alpha}}+\frac{1}{2}m_{\alpha}\omega_{\alpha}^{2}(x_{\alpha}+d_{\alpha})^{2} \right)$ where $\omega_{01}$ is the qubit excitation energy. The displacement $d_{\alpha}$ along the $\alpha$-th mode determines the strength of the qubit-bath interaction, as measured by the reorganization energy $\lambda_{\alpha}=\frac{m_{\alpha}\omega_{\alpha}^{2}d_{\alpha}^{2}}{2}$ . This becomes more transparent when the Hamiltonian is expressed in the form $H=H_{\text{S}}+H_{\text{B}}+H_{\text{SB}}$ for which
\begin{equation}
    H_{\text{SB}}=\op{1}{1}\otimes\sum_{\alpha}\sqrt{2m_{\alpha}\lambda_{\alpha}}\omega_{\alpha}x_{\alpha}.
\end{equation}

Inserting the bath operator $B_{\alpha}=\sqrt{2m_{\alpha}\lambda_{\alpha}}\omega_{\alpha}x_{\alpha}$ in \eq{eq::Decoherence-Function} we obtain the decoherence function
\begin{equation}\label{eq:decoherence-function-discrete}
    |\Phi_{01}(t)|= \exp \left \{ - \sum_{\alpha} \left [\lambda_{\alpha}\omega_{\alpha}  \coth{\left(\frac{\omega_{\alpha}} {2 k_{B}T}\right)}\frac{1- \cos (\omega_{\alpha} t)}{\omega_{\alpha}^2} \right ] \right\} 
\end{equation}
associated with a discrete  environment, where $k_{B}T$ represents the thermal energy. 

To generalize \eq{eq:decoherence-function-discrete} to the continuous limit, we introduce the spectral density\cite{Leggett1987} $J(\omega)= \sum_{\alpha} \lambda_{\alpha}\omega_{\alpha} \delta(\omega-\omega_{\alpha})$, a quantity that summarizes the frequencies of the bath and their coupling strength to the qubit. This allows us to express $|\Phi_{01}(t)|$ as: 

\begin{equation}\label{eq:decoherence-function-continous} |\Phi_{01}(t)| = \exp \left\{- \int_{0}^{\infty} d\omega\, J(\omega) \coth{\left(\frac{\omega} {2 k_{B}T}\right)}\frac{1- \cos (\omega t)}{{\omega^2}}\right \}. \end{equation}
Consequently, the loss of electronic coherence, is given by $\left | \frac{ [\tilde{\rho}_{\text{S}}]_{\text{01}}(t)}{ [\tilde{\rho}_{\text{S}}]_{\text{01}}(0)} \right | = |\Phi_{01}(t)|$. As is evident in \eq{eq:decoherence-function-continous}, the qubit decoherence does not follow a simple Gaussian or exponential decay. In turn, it is determined by the exponential of oscillatory functions, with periods determined by the bath's  frequencies.

\subsection{Requirements for strict exponential decay }\label{sec:Exponential}
 
When is the decoherence strictly exponential? Consider first the case when there is only a finite number of modes, $\alpha$, whose decoherence function is described by Eq.~\eqref{eq:decoherence-function-discrete}.
In this case, the decoherence function is expected to have a Poincar\'e recurrence time of $2\tau$, which can be long or short depending on the bath. 
We thus suppose that the decoherence function has a periodicity of $2\tau$, and impose the condition that the coherence decays exponentially at initial times. That is, $\ln |\Phi_{01} (t)| =-\gamma |t|$, for $-\tau< t <\tau$ and $\ln \Phi_{01}(t+2\tau) =\ln |\Phi_{01} (t)| $ for all $t$. Here, $\gamma > 0$ is a real constant that quantifies the exponential decoherence rate. 

Since $\ln |\Phi_{01} (t)| $ is an even function and has a period of $2\tau$, the Fourier series of $\ln |\Phi_{01} (t)| $ will consist only of cosine terms: 
\begin{equation} 
\ln |\Phi_{01} (t)|  =  \frac{a_{0}}{2} + \sum_{n=1}^{+\infty} a_n \cos \left(\frac{n\omega_0 t}{2}\right) ,
\end{equation} 
where $\omega_{0}= {2\pi}/{\tau}$ is the fundamental frequency that dictates the overall recurrence time. The Fourier coefficients are given by: 
\begin{align} 
a_n &= \frac{1}{\tau} \int_{-\tau}^{\tau} |\Phi_{01} (t)| \cos\left(\frac{n \omega_0 t}{2}\right) dt \nonumber \\
&= \frac{4 (1-(-1)^n) \gamma}{\pi n^2\omega_0} \quad \text{for} \quad n > 0, 
\end{align} 
and
\begin{equation}
    a_0 = \frac{1}{\tau} \int_{-\tau}^{\tau}|\Phi_{01} (t)|dt = - \frac{2\pi\gamma}{\omega_0}.
\end{equation}
This leads to:
\begin{equation} \label{eq:exp-discrete}
\ln|\Phi_{01} (t)| =- \frac{\pi\gamma}{\omega_0} + \frac{2\gamma\omega_0}{\pi}\sum_{n=1}^{+\infty} \frac{\cos\left((n-\frac{1}{2})\omega_0 t\right)}{\left((n-\frac{1}{2})\omega_0\right)^2}.
\end{equation} 

By comparing \eq{eq:decoherence-function-discrete} and \eq{eq:exp-discrete}, we observe that to achieve purely exponential decay, one requires: 
\begin{subequations} \label{eq:finite-temp}
\begin{align} \omega_{\alpha} &= \left(\alpha-\frac{1}{2} \right)\omega_0, \quad \alpha = 1,\ 2,\ \ldots
\\ 
%\lambda_{\alpha}\omega_{\alpha}&= \gamma\omega_0
\lambda_{\alpha} &= \frac{2\gamma}{\pi\left(\alpha - \frac{1}{2}\right)\coth\left(\omega_{\alpha}/(2k_{B}T)\right)},
\end{align} 
\end{subequations}
where we have used $\sum_{\alpha=1}^{\infty} - 2\gamma /(\pi (\alpha-1/2)^{2}\omega_{0})=-\pi \gamma/\omega_{0}$.
That is, for purely exponential decay all temperature-weighted qubit-bath couplings $\lambda_\alpha \omega_\alpha \coth\left(\omega_{\alpha}/(2k_{B}T)\right)$ need to be the same, and all frequencies need to be evenly spread and span all frequencies.

The zero-temperature case can be obtained in the $\omega_{\alpha}/(2k_{B}T) \to +\infty$ limit of Eq.~\eqref{eq:finite-temp} 
\begin{equation}\label{eq:const-g}
\lambda_{\alpha}=  \frac{2\gamma}{\pi\left(\alpha-\frac{1}{2}\right)}. 
\end{equation}
 In this limit, purely exponential decay emerges when $\lambda_{\alpha}\omega_{\alpha}$ are all the same, and evenly distributed across all frequencies.
 
In turn, for high temperatures $k_{B}T \gg \omega_{\alpha}$ and $\coth (\omega_{\alpha}/(2k_{B}T))\approx2k_{B}T/\omega_{\alpha}$. In this limit, to achieve a purely exponential decay it is necessary that
\begin{equation}\label{eq:const-re}
\lambda_{\alpha}= \frac{\gamma\omega_0}{\pi k_{B}T}. 
\end{equation}
This implies that equally spaced frequencies and constant reorganization energy are required at the high-temperature limit. Thus, the requirements for strict exponential decoherence to emerge are very specific and thus rare in actual baths at zero, finite, and infinite temperatures.

We now extend this discussion to the continuous limit with $\tau \to \infty$ and $\omega_0 \to 0$. To obtain the strict exponential decay, we need: (i) a constant spectral density $J(\omega)$ at the zero temperature limit, or (ii) a purely Ohmic spectral density, $J(\omega)=\eta\omega$, with $\eta$ being a constant, at the high-temperature limit.
The later requirement is unphysical as it leads to infinite overall reorganization energy $\lambda = \int_{0}^{\infty} d\omega\,J(\omega)/\omega $.

To avoid the infinite overall reorganization energy, we next discuss the case with a cut-off frequency $\omega_{c}$ for the spectral density $J(\omega)$ such that for $\omega \to \infty$, $J(\omega) \to 0$. In this way, the overall reorganization energy is finite. In this case, the exponential behavior for the continuous limit Eq.~\eqref{eq:decoherence-function-continous} becomes apparent as time approaches infinity ($t \to \infty$). To see this, we express $\frac{1-\cos(\omega t)}{\omega^{2}}=\frac{1}{2} \frac{\sin^{2}( {\omega t}/{2})}{({\omega}/{2})^{2}}$, which converges to $\pi t \delta(\omega)$ as $\omega t \to \infty$ for $0 < \omega < \omega_{c}$. Consequently,
\begin{equation}\label{eq:exponential-limit}
\begin{aligned}
    |\Phi_{01}(t)| &= \exp \left \{ -\pi t \int_{0}^{\infty} d\omega\, J (\omega) \coth \left(\frac{\omega}{2k_{B}T}\right) \delta(\omega)\right\}
\\
    &= \exp \left\{- \pi t \lim_{\omega\to0^{+}}\left [ J (\omega) \coth \left(\frac{\omega}{2k_{B}T}\right) \right]\right\}
\end{aligned}
\end{equation}
Then, an exponential decay dominates in the long-time limit if $\lim_{\omega \to 0^{+}} J_{\alpha}(\omega) \coth \left(\frac{\omega}{2k_{B}T}\right)$ is finite and non-zero. 

Two Ohmic models of the spectral density are predominantly used in the literature, namely the Ohmic with exponential cut-off, $J_{\text{EXP}}(\omega)$, and the Ohmic with Lorentzian cut-off (Drude-Lorentz), $J_{\text{DL}}(\omega)$. They are defined as
\begin{subequations}\label{eq:dynamics}
    \begin{align}
    J_{\text{EXP}} (\omega)&=\frac{2}{\pi}\frac{\lambda\omega}{\omega_{c}}  e^{-\omega/\omega_{c}},\\
    J_{\text{DL}} (\omega)&=\frac{2}{\pi}\lambda \omega \frac{\omega_{c}}{\omega^{2}+\omega_{c}^{2}}.
 \end{align}
\end{subequations}
These two spectral densities yield a purely Ohmic form when $\omega \ll \omega_{c}$. In this range, $J(\omega)=2\lambda \omega / (\pi \omega_{c})$ and the decoherence function for long times Eq.~\eqref{eq:exponential-limit} becomes $|\Phi_{01}(t)| = e^{-rt}$ where $r=4k_B T\lambda /\omega_{c}$. Thus, exponential decoherence can be obtained when $ t  \gg 1/\omega_{c}$.

\begin{figure}[htb]
\centering
\includegraphics[width=0.4\textwidth]{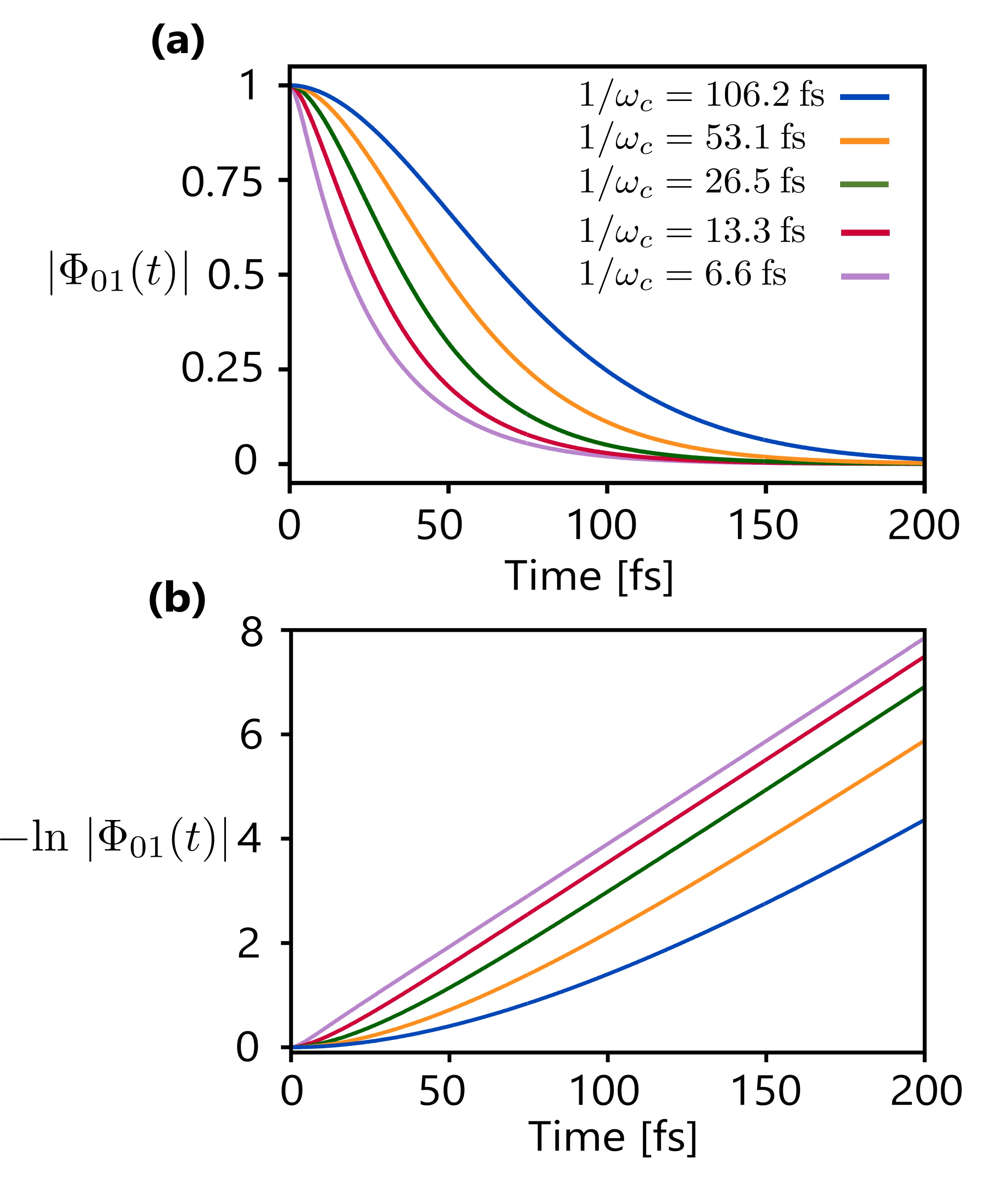}
\caption{\textbf{Decoherence due to an Ohmic bath with varying cut-off frequency $\omega_{c}$.} (a) Decoherence function and (b) its natural logarithm with  $T=300~\mathrm{K}$ and $\lambda=0.5\:\omega_{c}$. Note how the decoherence becomes purely exponential as $t  \gg 1/\omega_{c}$.
}
\label{fig:exponential}
\end{figure}

To numerically illustrate these observations, \fig{fig:exponential} shows the decoherence dynamics induced by a Drude-Lorentz bath [\eq{eq:dynamics}b] for varying cut-off frequencies. \Fig{fig:exponential}a depicts the decoherence function, and \fig{fig:exponential}b its natural logarithm. We fix the ratio $\lambda/\omega_{c} =0.5$ to have the same exponential decay rate for different $\omega_{c}$. As the cut-off frequency increases, the decoherence function begins to exhibit exponential decay more rapidly. This is more distinctly observed as a linear trend in the natural logarithm. For example, with $1/\omega_{c} = 106.2~\mathrm{fs}$ (blue line), the exponential regime only becomes dominant for $t \sim 150$ fs, when almost all electronic coherence has already been lost. By contrast, at $1/\omega_{c} = 26.5~\mathrm{fs}$ (green line), the exponential regime appears after $\sim 35$ femtoseconds, when the electronic coherence is around 0.4. As the cut-off frequency further increases to $1/\omega_{c} = 6.6~\mathrm{fs}$ (purple line), the exponential regime starts to dominate even at initial times, where the decoherence function is approximately 1.

In summary, purely exponential coherence decay is expected at zero temperature for a constant spectral density and at the high-temperature limit for a purely Ohmic spectral density. At finite temperatures, the exponential regimes dominate at times longer than the bath correlation ($t \gg 1/\omega_{c}$). In the latter case, as the bath correlation time increases, it becomes increasingly difficult to observe the exponential regime as it arises when most of the coherence has been lost. We explore the relevance of the exponential regime in realistic chemical systems in Sec. \ref{sec:Results} E.

\subsection{Gaussian decay is always present at early times.  }\label{sec:Gaussian}
We now seek to isolate conditions that result in a purely Gaussian decay. To this end, our starting point is \eq{eq:decoherence-function-continous}. Expanding the cosine function using a Taylor series around $t= 0$ in time yields

\begin{figure}[htb]
\centering
\includegraphics[width=0.4\textwidth]{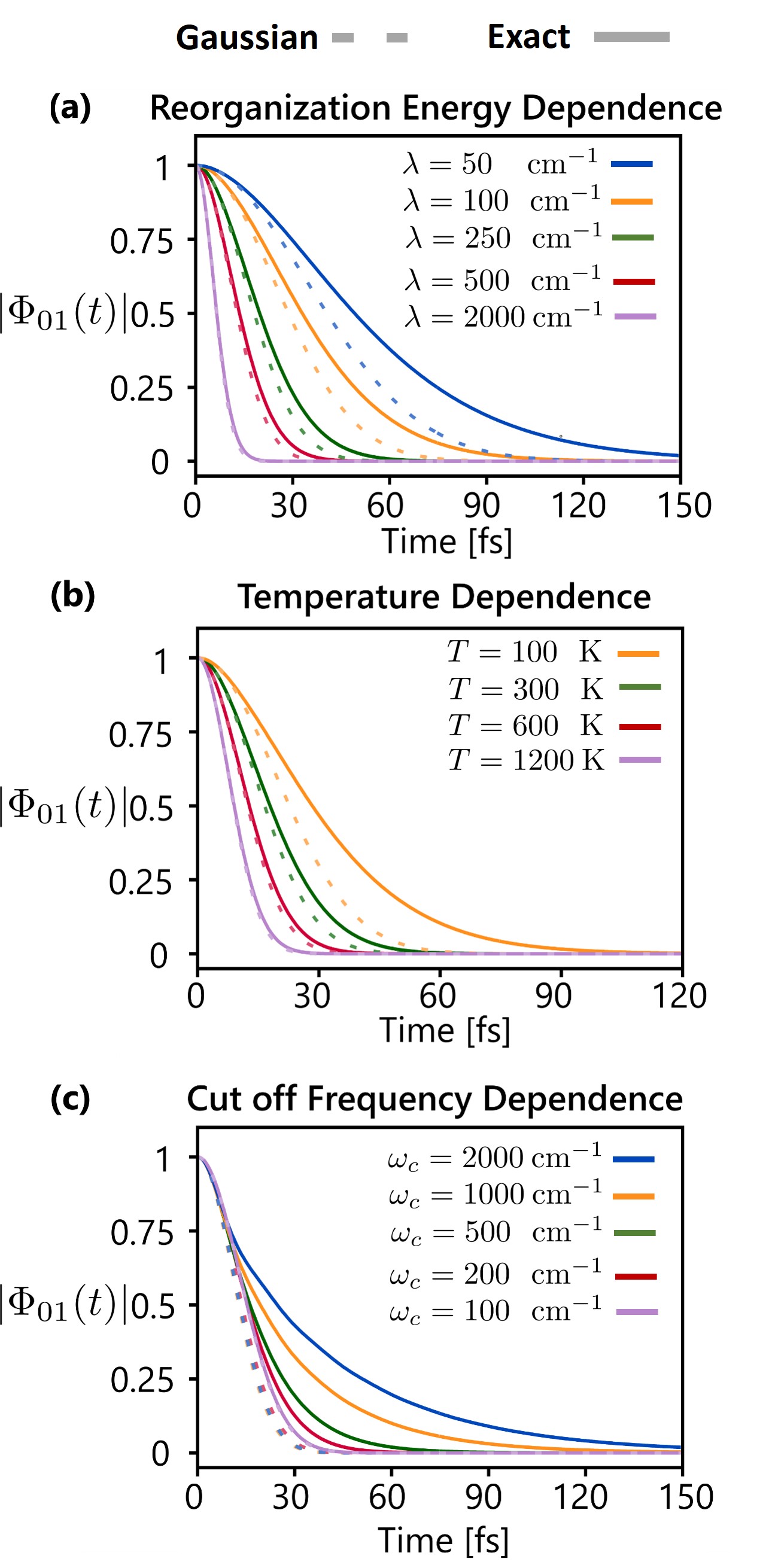}
\caption{\textbf{Relevance of the Gaussian regime as a function of:} (a) Reorganization energy (for $\omega_{c}=100~\mathrm{cm}^{-1}$ and $T=300~\mathrm{K}$), (b) temperature (for $\omega_{c}=100~\mathrm{cm}^{-1}$ and $\lambda=300~\mathrm{cm}^{-1}$), and (c) cut off frequency (for $\lambda=400~\mathrm{cm}^{-1}$ and $T=300~\mathrm{K}$) for a Drude-Lorentz spectral density. Dashed lines: short-time Gaussian approximation. Solid lines: exact with all higher orders included. As the reorganization energy and temperature increase and the bath correlation time decreases,  the Gaussian regime becomes more accurate because the coherence is lost more rapidly. }
\label{fig:Gaussian}
\end{figure}

\begin{equation}\label{eq:Gaussian}
\begin{aligned}
        |\Phi_{01}(t)| = \exp &\left\{ \int_{0}^{\infty} d\omega\, \frac{J(\omega)}{\omega^{2}} \coth{\left(\frac{\omega} {2 k_{B}T}\right)} \right .\\ 
        &\left .  \sum_{n=1}^{\infty} \frac{(-1)^{n}(\omega t)^{2n}}{2n!} \right \}.
\end{aligned}
\end{equation}
From this expression, it is clear that the Gaussian decay, arising for $n=1$, is universally present at early times for initially separable states (provided that the integral converges). 

This initial Gaussian decay is well-known, what remain unclear is to understand when is it dominant and its degree of applicability in molecular qubits. To obtain conditions that yield a purely Gaussian decay for all times, we numerically evaluate the importance of higher order terms, $n > 1$, in \eq{eq:Gaussian} as a function of the reorganization energy, temperature, and the cut-off frequency.

The validity of the Gaussian region for a Drude-Lorentz spectral density is shown in \Fig{fig:Gaussian}. It is clear that increasing the reorganization energy (\fig{fig:Gaussian}a) and the temperature (\fig{fig:Gaussian}b) improves the validity of the Gaussian regime. This is because increasing these parameters makes the decoherence decay faster and the early time approximation more accurate. By contrast, as the cut-off frequency is increased (\fig{fig:Gaussian}c), the accuracy of the Gaussian approximation deteriorates as the decoherence decays slower, leading to a smaller validity range for the short-time expansion.

That is, Ohmic baths with short correlation time (or high cut-off frequencies) yield exponential decay, while those with long correlation times yield Gaussian decay. 

\subsection{Decoherence and lineshapes in spectroscopy}\label{sec:spectroscopy}

Decoherence is closely tied to the theory of spectroscopic lineshapes. In fact, it is known that electronic temporal decoherence patterns and time scales can be estimated by performing a Fourier transform on the lineshapes observed in electronic absorption and fluorescence spectra in the pure dephasing limit.\cite{schatz2002quantum,tokmakotime} This is because for a Hamiltonian in the form of \eq{eq:Molecular-Hamiltonian}, and under the Born-Oppenheimer and Condon approximation, the absorption and fluorescence lineshapes are given by  
\begin{equation}\label{eq:abs}
    \sigma_{\text{A}}(\omega)=\frac{1}{\pi} \text{Re}\left\{ \int_{0}^{\infty}dt \, \exp \left \{ i(\omega-\omega_{{01}})t \right\} \exp \left\{-g(t) \right\} \right\}
\end{equation}
\begin{equation}\label{eq:flu}
    \sigma_{\text{F}}(\omega)=\frac{1}{\pi} \text{Re}\left\{ \int_{0}^{\infty}dt \, \exp \left \{ i(\omega-\omega_{{01}}+2\lambda)t \right\} \exp \left\{-g^{*}(t) \right\} \right\}
\end{equation}
where $\omega_{{01}}$ is the electronic energy gap, $2\lambda$  the Stokes shift, and 
\begin{equation} \label{eq:linebroadening}
\begin{aligned}
    g(t)&= \int_{0}^{\infty} d\omega\, J(\omega) \coth{\left(\frac{\omega} {2 k_{B}T}\right)}\frac{1- \cos (\omega t)}{{\omega^2}}\\
    &+i\int_{0}^{\infty} d\omega\, J(\omega) \frac{\sin (\omega t)-\omega t}{{\omega^2}}.
    \end{aligned}
\end{equation}
the line-broadening function. Note that the real part of $g(t)$ is related to the decoherence function as $|\Phi_{01}(t)|=\exp \left \{ - \text{Re} \{ g(t)\} \right\}$. Thus, by Fourier transforming $\sigma_{\text{A}}(\omega)$ or $\sigma_{\text{F}}(\omega)$ one can extract $g(t)$ and thus the decoherence function.

By inspecting the lineshape of an absorption or fluorescence spectra, it is possible to determine if the decoherence is Gaussian, exponential, or neither of them. To show this, we apply the short-time approximation to \eq{eq:linebroadening} to get 
\begin{equation} \label{eq:st-approx}
g(t)=\frac{t^{2}}{2}\int_{0}^{\infty} d\omega\, J(\omega) \coth{\left(\frac{\omega} {2 k_{B}T}\right)},
\end{equation}
as the imaginary term vanishes up to $\mathcal{O}(t^{3})$. Introducing \eq{eq:st-approx} into \eqs{eq:abs}{eq:flu} 
\begin{equation}\label{eq:FT-abs}
\begin{aligned}
    \sigma_{\text{A}}(\omega)&=\frac{1}{\pi} \text{Re}\left\{ \int_{0}^{\infty}dt \, \exp \left \{ i(\omega-\omega_{{01}})t \right\} |\Phi_{01}(t)| \right\}\\
    &=\sqrt{\frac{1}{2\pi \mean{\omega^{2}}}}\exp \left \{ - \frac{1}{2} \frac{(\omega-\omega_{{01}})^{2}}{\mean{\omega^{2}}} \right \},
\end{aligned}
\end{equation}
\begin{equation}\label{eq:FT-flu}
\begin{aligned}
    \sigma_{\text{F}}(\omega)&=\frac{1}{\pi} \text{Re}\left\{ \int_{0}^{\infty}dt \, \exp \left \{ i(\omega-\omega_{{01}}+2\lambda)t \right\} |\Phi_{01}(t)| \right\}\\
    &=\sqrt{\frac{1}{2\pi \mean{\omega^{2}}}}\exp \left \{ - \frac{1}{2} \frac{(\omega-\omega_{{01}}+2\lambda)^{2}}{\mean{\omega^{2}}} \right \}
\end{aligned}
\end{equation}
where we have defined $\mean{\omega^{2}}=\int_{0}^{\infty} d\omega\, J(\omega) \coth{\left(\frac{\omega} {2 k_{B}T}\right)}$. 
Note that the Gaussian limit in the decoherence function produces an absorption and fluorescence lineshape that is Gaussian and identical but separated by a Stokes shift given by $2\lambda$ (as long as $\mean{\omega^{2}}$ converges). From a quantum perspective this Gaussian limit arises from entanglement between the system and the bath.

To obtain Lorentzian-like lineshapes, following Skinner\cite{skinner1986pure,hsu1984thermal}, we let $\omega t\to\infty$ in the line-broadening function Eq. \eqref{eq:linebroadening} to get
\begin{equation}\label{eq:lt-limit}
\begin{aligned}
    g(t) &=  \pi t \lim_{\omega\to0^{+}}\left [ J (\omega) \coth \left(\frac{\omega}{2k_{B}T}\right) \right] \\
    & -i t \int_{0}^{\infty} d\omega\,  \frac{J(\omega)}{{\omega}}\\
    & = t \mathcal{C}^{0} - i \lambda t
\end{aligned}
\end{equation}
Here $\mathcal{C}^{0}= \pi \lim_{\omega\to0^{+}}\left [ J (\omega) \coth \left(\frac{\omega}{2k_{B}T}\right) \right]$. Introducing \eq{eq:lt-limit} into \eqs{eq:abs}{eq:flu}
\begin{equation}
    \sigma_{\text{A}}(\omega)=\sigma_{\text{F}}(\omega)=\frac{1}{\pi}\frac{\mathcal{C}^{0}}{(\omega-\omega_{{01}})^{2}+(\mathcal{C}^{0})^{2}}.
\end{equation}
In this case, the Stokes shift between absorption and fluorescence vanishes as the bath motion is very fast such that the radiation field observes an averaged two-level system (as long $\mathcal{C}^{0}$ converges).\cite{Mukamel_book} Thus, an exponential decay of coherence leads to Lorentzian lineshapes.

In general, pure dephasing processes can be mimicked by classical noise.\cite{Gu_2019} In the Gaussian limit, which gives Gaussian lineshapes, this is achieved by introducing static noise in the initial conditions, which leads to the interpretation that this is an inhomogeneous process. By contrast, exponential decoherence decay, which produces Lorentzian lineshapes and is commonly known as the homogeneous limit, requires colored noise.\cite{Gu_2019,skinner1986pure} This distinction was first recognized through Kubo's stochastic model\cite{kubo1962,kubo1969}, see 
 Ref.\footnote{Classical noise processes have been extensively used to model line-broadening functions in spectroscopy. \cite{Mukamel_book,kubo1962,kubo1969,tokmakotime} In this model, the effect of the bath is incorporated by letting the electronic energy gap, $\omega_{01}$, fluctuate in time $\omega_{01} \to \omega_{01} + \delta\omega_{01}(t)$, where $\delta\omega_{01}(t)$ is a random function of time given by a Gaussian distribution. The resulting classical frequency correlation function is given by 
\begin{equation}
\begin{aligned}
    {C_{\delta\omega_{01}}(t)}&{=\left \langle \delta\omega_{01}(t) \delta\omega_{01}(0) \right\rangle} \\
    &{=\Delta^{2} \exp \left [-t/\tau_{c} \right].}
\end{aligned}
\end{equation} 
Here, $\Delta = \sqrt{\left \langle \delta \omega_{01}^{2}\right \rangle}$ is the variance of the Gaussian distribution of frequencies and $\tau_{c}$ is the correlation time. The resulting, classical $g_{c}(t)$, line-broadening function is\cite{Mukamel_book,tokmakotime} 
\begin{equation}
    \begin{aligned}
         {g_{c}(t)} & {= - \int_{0}^{t}d\tau \, (t-\tau)  \left \langle \delta\omega_{01}(\tau) \delta\omega_{01}(0) \right\rangle} \\
        &{= \Delta^{2} \tau_{c}^{2} \left ( \exp(-t/\tau_{c})+t/\tau_{c}-1 \right ).}
    \end{aligned}
\end{equation}  
It is important to note that the classical line-broadening function is purely real. Thus, it does not capture Stokes shifts and does not satisfy the fluctuation-dissipation theorem.\cite{Mukamel_book}
In the limit where $t \ll \tau_{c}$ the classical correlation function becomes a constant, $C_{\delta\omega_{01}}(t)\approx\Delta ^{2}$, and the line-broadening function yields $g_{c}(t)=\Delta^{2}t^{2}/2$. Because the correlation function is a constant, this is known as the inhomogeneous limit. Note that if $\Delta^{2}=\int_{0}^{\infty} d\omega\, J(\omega) \coth{\left(\frac{\omega} {2k_{B}T}\right)}$ we retrieve \eq{eq:st-approx}. In turn, in the homogeneous limit, we have that $t \gg \tau_{c}$ and the classical line-broadening function becomes $g_{c}(t)=\Delta^{2}\tau_{c}t$. Thus, we can mimic the real part of \eq{eq:lt-limit} by setting $\Delta^{2}\tau_{c} = \mathcal{C}^{0}$, but we are completely missing the imaginary component. However, this does not mean that quantum and classical noise processes are equivalent but rather that in this limit classical noise processes can mimic quantum decoherence because the line broadening function is purely real, as previously noted in the literature.\cite{skinner1986pure,Bing2017,Vezvaee2024}.} for a summary.

However, even when classical noise processes can effectively mimic decoherence behavior in the pure dephasing limit, from a quantum perspective, the decoherence arises from system-bath entanglement and not from classical noise. The influence of inhomogeneities in the initial condition can be effectively distinguished from decoherence through spin-echo experiments\cite{nian2024spin,cucchietti2005,zurek2007echo} where a filter can eliminate the influence of classical noise in the initial conditions but not quantum decoherence due to system-bath entanglement. The consequence of this observation is that Gaussian spectral peaks, usually associated with inhomogeneous broadening, can emerge from quantum decoherence due to the entangling unitary system-bath dynamics, even when there is no inhomogeneity in the initial conditions. 

\subsection{Decoherence using realistic spectral densities}\label{sec:real-molecule}

\begin{figure}[htb]
\centering
\includegraphics[width=0.45\textwidth]{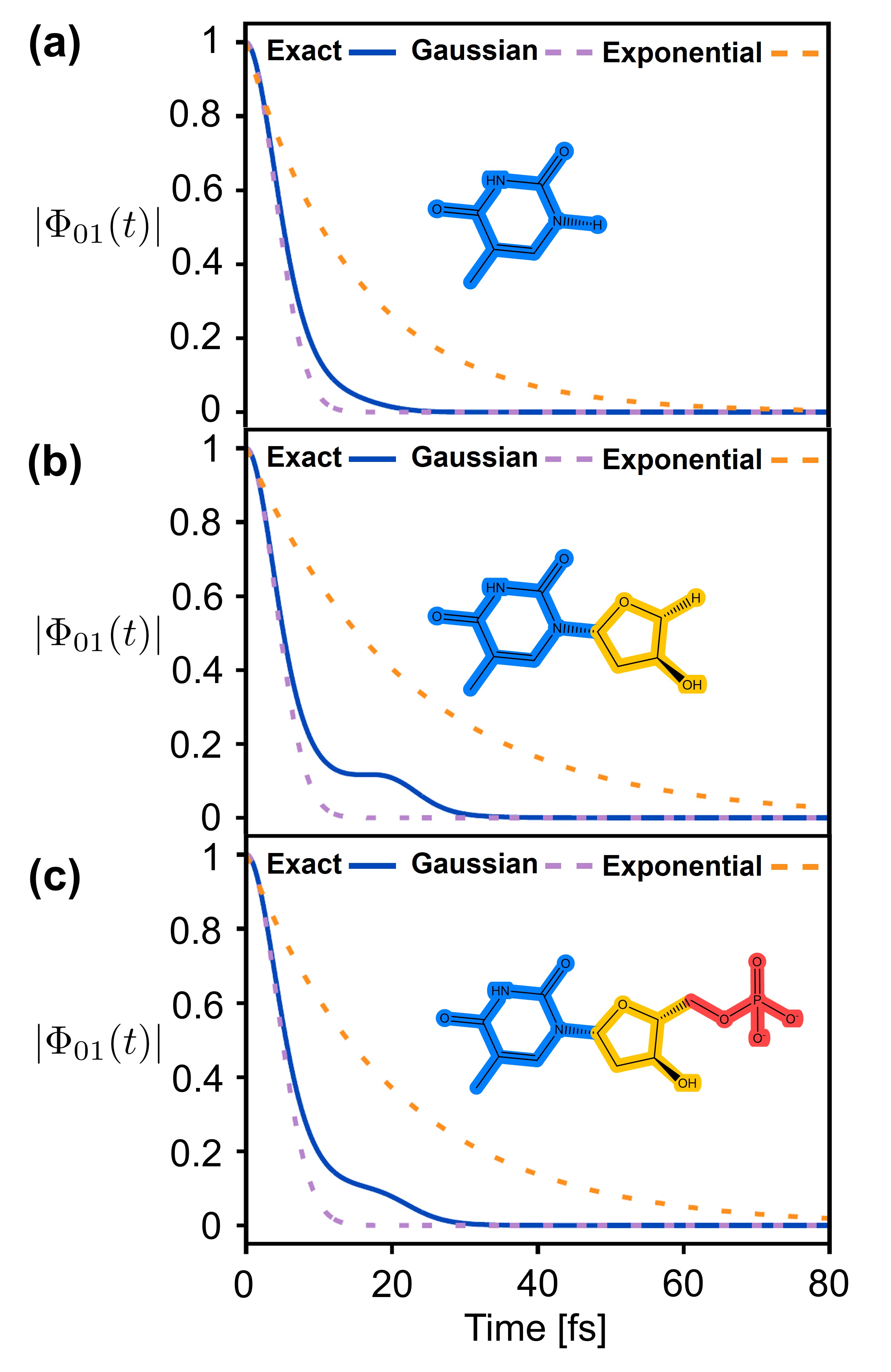}
 \caption{\textbf{Electronic decoherence dynamic in (a) thymine, (b) its nucleoside, and (c) its nucleotide.} Solid lines are the decoherence dynamics reconstructed from spectral densities extracted from experiments.\cite{Gustin2023,yarasi2007,billinghurst2012} The Gaussian initial decay is shown in purple, and the exponential in orange. The parameters used in this simulation can be found in the supplementary information of Ref. \cite{Gustin2023}} 
\label{fig:molecules}
\end{figure}
We now examine the relevance of the Gaussian and exponential coherence decay models for electronic decoherence in realistic molecules immersed in condensed phase baths. This is important since the Gaussian and exponential decoherence time scale are often employed to quantify electronic decoherence in molecular systems.\cite{schatz2002quantum,shu2023}

In the condensed phase, the spectral density of molecules has a wide low-frequency component that accounts for interactions with solvent modes. In addition, a series of $N$ sharp peaks capture the influence of intramolecular vibrations. The total spectral density can be expressed as:
\begin{equation}
    J(\omega)=\left (\sum_{\alpha=0}^{N} J_{\alpha}(\omega) \right ) ,
\end{equation}
where the solvent is represented through a Drude-Lorentz model
\begin{equation}
\label{eq:DL-SPD}
J_{0}(\omega)=\frac{2}{\pi}\lambda_{0}\frac{\omega\omega_{c}}{\omega^{2}+\omega_{c}^{2}}
\end{equation}
while the contribution of the $\alpha$-th ($\alpha \geq 1$) vibrational modes is modeled through a Brownian oscillator spectral density
\begin{equation}\label{eq:B-SPD}
    J_{\alpha}(\omega) =\frac{2}{\pi}\lambda_{\alpha}\omega^{2}_{\alpha}\frac{\omega \gamma_{\alpha}}{\left(\omega^{2}_{\alpha}-\omega^{2}\right)^{2}+\omega^{2}\gamma^{2}_{\alpha}}, \quad (\alpha\geq1).
\end{equation}
Here, $\lambda_{\alpha}$ is the reorganization energy of the solvent ($\alpha=0$) or the $\alpha$-th vibrational mode ($\alpha \neq 0$), ${\gamma_{\alpha}}$ are the vibrational lifetimes and $1/\omega_{c}$ is the solvent correlation time. The decoherence function in this case yields
\begin{equation}\label{eq:DL-real}
\begin{split}
        |\Phi_{01}(t)| = \exp \left\{ \sum_{\alpha=0}^{N} \int_{0}^{\infty} d\omega\, J_{\alpha}(\omega) \coth{\left(\frac{\omega}{2 k_{B}T}\right)} \right. \\
        \left. \frac{1 - \cos (\omega t)}{\omega^2} \right\}.
\end{split}
\end{equation}

The Gaussian regime can be achieved when the short-time limit is considered, such that $1-\cos(\omega t)\approx\frac{\omega^{2}t^{2}}{2}$. This results in \begin{equation}  
|\Phi_{01}(t)|=\exp \left\{\sum_{\alpha=0}^{N} \frac{t^{2}}{2}\int_{0}^{\infty} d\omega\, J_{\alpha}(\omega) \coth{\left(\frac{\omega} {2 k_{B}T}\right)} \right\} 
\end{equation} 
By contrast, the exponential behavior becomes important for long times ($t \to \infty$), see \eq{eq:exponential-limit}. For the spectral density in \eqs{eq:DL-SPD}{eq:B-SPD}, such exponential decay becomes
\begin{equation}
   |\Phi_{01}(t)|= \exp \left\{ -4 k_{\text{B}}Tt \left ( \frac{\lambda_{0}}{\omega_{c}}+\sum_{\alpha=1}^{N} \frac{\gamma_{\alpha}\lambda_{\alpha}}{\omega_{\alpha}^{2}} \right)\right \}.
\end{equation}

While the theoretical framework suggests that both a Gaussian and exponential regime always emerge in electronic decoherence, their quantitative importance in realistic molecular qubits remains unclear. To evaluate this, \fig{fig:molecules} compares the exponential and Gaussian decay and the overall decoherence dynamics of thymine, its nucleoside, and nucleotide immersed in 300 K water. The spectral densities for these molecules were recently reconstructed from resonance Raman spectroscopy\cite{Gustin2023}, opening the opportunity to investigate its decoherence with realistic complexity. 
As shown, the exponential regime is only relevant when all electronic coherence has already been lost. In turn, the Gaussian decay accurately reproduces the overall decoherence dynamics at initial times but fails to capture coherence recurrences and overestimates the overall decoherence by a factor of $\sim 2$ (ratio of the time need for the exact and Gaussian approximation to reach $|\Phi_{01}(t)|=0.01$). Thus, the Gaussian decoherence model is a more appropriate (albeit imperfect) description of electronic coherence dynamics for molecules in condensed phase baths. 

We note that in thymine, \fig{fig:molecules}a, the decoherence profile does not show recurrences. These recurrences arise from the wavepacket evolution of intramolecular vibrations. In this case, the visibility of the effect is suppressed because the solvent-induced decoherence is accelerated by the additional hydrogen bond of thymine in water with respect to its nucleoside and nucleotide.\cite{Gustin2023} 

\subsection{Influence of initial entanglement on quantum decoherence  }\label{sec:entangled}

In decoherence studies, it is customary to use initially separable states. However, in nature, these
separable states are exotic and difficult to prepare. For example, for electronic decoherence, these states arise only in the limit where the Born-Oppenheimer approximation is exact and, as such, represent an idealization.\cite{izmaylov2017entanglement} While in some situations, these separable states can be a useful approximation to the true eigenstates (e.g., when the diabatic electronic ground state is effectively decoupled from higher-lying electronic states), strictly speaking, they are rare.\cite{Marceau2019,Chang2024} Further, the preparation of superposition states using laser pulses can also lead to additional electron-nuclear entanglement when non-Condon effects are important for impulsive excitation or when the photoexcitation is not in the impulsive limit.\cite{Wen_2020} Thus initial entanglement is common in molecular systems. How does this initial qubit-bath entanglement influence the decoherence dynamics?

A partial answer to this question was offered at zero temperature for electronic decoherence, where the decoherence arises due to nuclear wavepacket evolution in alternative electronic potential energy surfaces (PES).\cite{jasper2005electronic,jasper2006non} In this context, it was shown that if the initial nuclear wavepackets in the two surfaces involved coincide spatially but differ in initial momentum (and thus represent an initially entangled state), then the initial time decoherence transitions from Gaussian to exponential for early times. 

To understand this problem in a more general framework, we consider the qubit and bath to be entangled as:
\begin{equation}\label{eq::entangled}
\ket{\Psi (0)}= c_{0}\ket{0}\ket{\chi_{0}} + c_{1}\ket{1}\ket{\chi_{1}}.
\end{equation} 
Here $\ket{\chi_{\text{n}}}$  is the bath wavepacket evolving in the ground $n = 0$ or excited $n =1$ qubit state. To relate $\ket{\chi_{1}}$ and $\ket{\chi_{0}}$ 
we introduce the displacement operator
\begin{equation}
    \hat{D}_\alpha(z )=\exp(z a_\alpha^{\dagger}- z^* {a_\alpha}),
\end{equation}
where $a_\alpha = \sqrt{\frac{m_\alpha \omega_\alpha}{2}} x_\alpha + \frac{i}{\sqrt{2 m_\alpha \omega_\alpha}} p_\alpha$.
This displacement operator reduces to the position translation operator when $z$ is real 
$\hat{D}_\alpha(z) = \hat{T}_{x_\alpha\to x_\alpha+\sqrt{2 / ( m_{\alpha }\omega_{\alpha})}z}$, 
and to the momentum translation operator when $z$ is purely imaginary 
$\hat{D}_\alpha(z) = \hat{T}_{p_\alpha\to p_\alpha+\sqrt{2  m_{\alpha }\omega_{\alpha}}z/i}$.
In general, to create an initial entanglement, we let
\begin{equation}
\ket{\chi_{1}}=\prod_{\alpha} 
\hat{D}((r_{\alpha}+ i s_{\alpha})/\sqrt{2})
\ket{\chi_{0}}.
\end{equation} 
That is, the wavepacket on the excited state is displaced by some momentum $s_\alpha \sqrt{m_{\alpha}\omega_{\alpha}}$ and position $r_\alpha/\sqrt{m_{\alpha}\omega_{\alpha}}$ for mode $\alpha$ with respect to $\ket{\chi_{0}}$.
The quantum decoherence function, $|\Phi_{01}(t)|=| \langle \chi_{0}(t) | \chi_{1}(t) \rangle |$, for this initially entangled state can be obtained by using the same steps described in \stn{sec:decoherence-functions} to get 

\begin{multline} \label{eq:entangled} 
\ln |\Phi_{01}(t)| = -\sum_{\alpha}  \Bigg[\frac{1}{4} \left(s_\alpha^2 + r_\alpha^2\right) \\+
\left({\frac{\lambda_{\alpha}}{\omega_{\alpha}}} + r_\alpha\sqrt{\frac{\lambda_{\alpha}}{2\omega_{\alpha}}}\right)\big(1-\cos(\omega_\alpha t)\big) \\
- s_\alpha\sqrt{\frac{\lambda_{\alpha}}{2\omega_{\alpha}}} \sin(\omega_\alpha t) \Bigg]. 
\end{multline}

This equation can be extended to finite temperatures (see supplementary material) to yield.

\begin{multline} \label{eq:entangled} 
\ln |\Phi_{01}(t)| = -\sum_{\alpha} \coth\left( \frac{\omega_\alpha}{2k_B T} \right)  \Bigg[\frac{1}{4} \left(s_\alpha^2 + r_\alpha^2\right) \\+
\left({\frac{\lambda_{\alpha}}{\omega_{\alpha}}} + r_\alpha\sqrt{\frac{\lambda_{\alpha}}{2\omega_{\alpha}}}\right)\big(1-\cos(\omega_\alpha t)\big) \\
- s_\alpha\sqrt{\frac{\lambda_{\alpha}}{2\omega_{\alpha}}} \sin(\omega_\alpha t) \Bigg]. 
\end{multline}

The entanglement due to displacement in position ($r_{\alpha}\neq 0 $) retains the same functional form of \eq{eq:decoherence-function-discrete} and thus preserves the initial Gaussian decay. By contrast, the entanglement due to displacement in momentum ($s_{\alpha}\neq 0 $) introduces oscillatory terms in \eq{eq:entangled} proportional to $\sin (\omega_{\alpha}t)$ that break the initial Gaussian decay and introduce a term that decays linearly in time for initial times that are consistent with exponential decay. Specifically, for early times now  $\ln |\Phi_{01}(t)| = \ln |\Phi_{01}(0)| - \gamma_{1} t - \gamma_{2} t^2 - \mathcal{O}(t^3)$
\begin{align}
\gamma_{1} &= -\sum_{\alpha}  s_\alpha\sqrt{\frac{\lambda_{\alpha}}{2\omega_{\alpha}}}\coth\left( \frac{\omega_\alpha}{2k_B T} \right), \quad\text{and}\\
    \gamma_{2} &= \frac{1}{2}\sum_{\alpha}\left( {\frac{\lambda_{\alpha}}{\omega_{\alpha}}} + r_\alpha\sqrt{\frac{\lambda_{\alpha}}{2\omega_{\alpha}}} \right) \coth\left( \frac{\omega_\alpha}{2k_B T} \right).
\end{align}

If we neglect higher-order terms $|\Phi_{01}(t)|/|\Phi_{01}(0)|\approx e^{-\gamma_{1}t}e^{-\gamma_{2}t^{2}}$ and thus $\gamma_{1}$ lead to initial exponential decay and $\gamma_{2}$ to Gaussian decay. As seen, $\gamma_{1} \neq 0$ when $s_{\alpha}\neq 0$.
Increasing the initial entanglement due to displacement in momentum space makes the initial exponential part increasingly dominant.
By contrast, entanglement due to displacement in position ($r_{\alpha} \neq 0$) retains the Gaussian shape but modifies the decoherence rate. As the initial entanglement increases ($|r_\alpha|,|\ s_\alpha|$ increase), the initial time coherence is reduced to $|\Phi_{01}(0)| = \exp\left(-\frac{1}{4}\sum_\alpha  \left(s_\alpha^2 + r_\alpha^2\right)\coth\left( \frac{\omega_\alpha}{2k_B T} \right)\right) \leq 1$, with the equality holding for the initially unentangled case.
This reflects the decay of initial coherence due to qubit-bath entanglement.

In principle, \eq{eq:entangled} could be extended to the continuous limit by defining new spectral densities associated with $r_{\alpha}$ and $s_{\alpha}$. However, the precise form of these spectral densities remains unclear, as existing formulations in the literature have been developed exclusively for unentangled initial conditions. Furthermore, qubit-bath entanglement may emerge from interactions that are more complex than displacement in the position and momentum of the bath states. However, the model remains highly informative when these entanglements dominate the decoherence dynamics and serve to demonstrate that initial entanglement has an important effect on the decoherence dynamics.

\section{Conclusion} \label{conclusions}
In conclusion, we provided an in-depth analysis of decoherence dynamics in molecular and other\cite{Vezvaee2024,paz2017multiqubit,kwiatkowski2020influence,cywinski2008enhance} qubits caused by thermal bosonic baths. Our analysis reveals that, in general, decoherence is neither purely Gaussian nor exponential but rather the exponential of oscillatory functions with periods determined by the bath frequencies. As shown, for initially unentangled qubit-bath states or initial entanglement due to displacement in the position of the bath states, the Gaussian decay is always present at early times. We find that it becomes increasingly dominant with increasing temperature, molecule-bath interactions, and bath correlation time (the latter applies only to unentangled initial conditions). By contrast, we find that strict exponential decay arises in very restrictive models of the spectral density that we isolate. However, it becomes dominant for times longer than the bath correlation time or , as shown here, for early times as we increase the initial entanglement due to momentum displacement of the bath states. 

We examined the applicability of the Gaussian and exponential coherence decay models for initial separable states in realistic molecules, i.e., thymine derivative in water at 300K. We find that the exponential regime only becomes relevant once the molecules lose most of their quantum coherence, and thus, it is not a good decoherence model for these molecules. In turn, Gaussian decay accurately reproduces the coherence loss at initial times but overestimates the overall decoherence by a factor of $\sim 2$. Thus, the Gaussian decoherence model is more appropriate, although imperfect, to describe electronic decoherence in condensed phase baths. 

In addition, we revisited the rich literature in molecular spectroscopy to show that the presence of Gaussian-shaped peaks in absorption and emission spectroscopy, commonly referred to as the static or inhomogeneous limit, does not imply the absence of quantum entanglement as widely believed. While Kubo's work linked Gaussian spectroscopic lineshapes to classical noise processes, this viewpoint overlooks the fact that, from a quantum perspective, these lineshapes result from system-bath entanglement generated by the unitary evolution of the composite system. Therefore, even though classical noise processes can effectively mimic decoherence behavior in the pure dephasing limit, this mimicry does not negate the presence of underlying quantum entanglement.

\begin{acknowledgments}
    This material is based upon work supported by the U.S. Department of Energy, Office of Science, Office of Basic Energy Sciences, Quantum Information Science Research in Chemical Sciences, Geosciences, and Biosciences Program under Award Number DE-SC0025334.
\end{acknowledgments}

\bibliography{references}

\begin{thebibliography}{81}%
\makeatletter
\providecommand \@ifxundefined [1]{%
 \@ifx{#1\undefined}
}%
\providecommand \@ifnum [1]{%
 \ifnum #1\expandafter \@firstoftwo
 \else \expandafter \@secondoftwo
 \fi
}%
\providecommand \@ifx [1]{%
 \ifx #1\expandafter \@firstoftwo
 \else \expandafter \@secondoftwo
 \fi
}%
\providecommand \natexlab [1]{#1}%
\providecommand \enquote  [1]{``#1''}%
\providecommand \bibnamefont  [1]{#1}%
\providecommand \bibfnamefont [1]{#1}%
\providecommand \citenamefont [1]{#1}%
\providecommand \href@noop [0]{\@secondoftwo}%
\providecommand \href [0]{\begingroup \@sanitize@url \@href}%
\providecommand \@href[1]{\@@startlink{#1}\@@href}%
\providecommand \@@href[1]{\endgroup#1\@@endlink}%
\providecommand \@sanitize@url [0]{\catcode `\\12\catcode `\$12\catcode `\&12\catcode `\#12\catcode `\^12\catcode `\_12\catcode `\%12\relax}%
\providecommand \@@startlink[1]{}%
\providecommand \@@endlink[0]{}%
\providecommand \url  [0]{\begingroup\@sanitize@url \@url }%
\providecommand \@url [1]{\endgroup\@href {#1}{\urlprefix }}%
\providecommand \urlprefix  [0]{URL }%
\providecommand \Eprint [0]{\href }%
\providecommand \doibase [0]{https://doi.org/}%
\providecommand \selectlanguage [0]{\@gobble}%
\providecommand \bibinfo  [0]{\@secondoftwo}%
\providecommand \bibfield  [0]{\@secondoftwo}%
\providecommand \translation [1]{[#1]}%
\providecommand \BibitemOpen [0]{}%
\providecommand \bibitemStop [0]{}%
\providecommand \bibitemNoStop [0]{.\EOS\space}%
\providecommand \EOS [0]{\spacefactor3000\relax}%
\providecommand \BibitemShut  [1]{\csname bibitem#1\endcsname}%
\let\auto@bib@innerbib\@empty
%</preamble>
\bibitem [{\citenamefont {Wasielewski}\ \emph {et~al.}(2020)\citenamefont {Wasielewski}, \citenamefont {Forbes}, \citenamefont {Frank}, \citenamefont {Kowalski}, \citenamefont {Scholes}, \citenamefont {Yuen-Zhou}, \citenamefont {Baldo}, \citenamefont {Freedman}, \citenamefont {Goldsmith}, \citenamefont {Goodson}, \citenamefont {Kirk}, \citenamefont {McCusker}, \citenamefont {Ogilvie}, \citenamefont {Shultz}, \citenamefont {Stoll},\ and\ \citenamefont {Whaley}}]{Wasielewski_2020}%
  \BibitemOpen
  \bibfield  {author} {\bibinfo {author} {\bibfnamefont {M.~R.}\ \bibnamefont {Wasielewski}}, \bibinfo {author} {\bibfnamefont {M.~D.~E.}\ \bibnamefont {Forbes}}, \bibinfo {author} {\bibfnamefont {N.~L.}\ \bibnamefont {Frank}}, \bibinfo {author} {\bibfnamefont {K.}~\bibnamefont {Kowalski}}, \bibinfo {author} {\bibfnamefont {G.~D.}\ \bibnamefont {Scholes}}, \bibinfo {author} {\bibfnamefont {J.}~\bibnamefont {Yuen-Zhou}}, \bibinfo {author} {\bibfnamefont {M.~A.}\ \bibnamefont {Baldo}}, \bibinfo {author} {\bibfnamefont {D.~E.}\ \bibnamefont {Freedman}}, \bibinfo {author} {\bibfnamefont {R.~H.}\ \bibnamefont {Goldsmith}}, \bibinfo {author} {\bibfnamefont {T.}~\bibnamefont {Goodson}}, \bibinfo {author} {\bibfnamefont {M.~L.}\ \bibnamefont {Kirk}}, \bibinfo {author} {\bibfnamefont {J.~K.}\ \bibnamefont {McCusker}}, \bibinfo {author} {\bibfnamefont {J.~P.}\ \bibnamefont {Ogilvie}}, \bibinfo {author} {\bibfnamefont {D.~A.}\ \bibnamefont {Shultz}}, \bibinfo {author} {\bibfnamefont {S.}~\bibnamefont {Stoll}},\ and\
  \bibinfo {author} {\bibfnamefont {K.~B.}\ \bibnamefont {Whaley}},\ }\bibfield  {title} {\bibinfo {title} {Exploiting chemistry and molecular systems for quantum information science},\ }\href {https://doi.org/10.1038/s41570-020-0200-5} {\bibfield  {journal} {\bibinfo  {journal} {Nat. Rev. Chem.}\ }\textbf {\bibinfo {volume} {4}},\ \bibinfo {pages} {490} (\bibinfo {year} {2020})}\BibitemShut {NoStop}%
\bibitem [{\citenamefont {McArdle}\ \emph {et~al.}(2020)\citenamefont {McArdle}, \citenamefont {Endo}, \citenamefont {Aspuru-Guzik}, \citenamefont {Benjamin},\ and\ \citenamefont {Yuan}}]{McArdle2020}%
  \BibitemOpen
  \bibfield  {author} {\bibinfo {author} {\bibfnamefont {S.}~\bibnamefont {McArdle}}, \bibinfo {author} {\bibfnamefont {S.}~\bibnamefont {Endo}}, \bibinfo {author} {\bibfnamefont {A.}~\bibnamefont {Aspuru-Guzik}}, \bibinfo {author} {\bibfnamefont {S.~C.}\ \bibnamefont {Benjamin}},\ and\ \bibinfo {author} {\bibfnamefont {X.}~\bibnamefont {Yuan}},\ }\bibfield  {title} {\bibinfo {title} {Quantum computational chemistry},\ }\href {https://doi.org/10.1103/RevModPhys.92.015003} {\bibfield  {journal} {\bibinfo  {journal} {Rev. Mod. Phys.}\ }\textbf {\bibinfo {volume} {92}},\ \bibinfo {pages} {015003} (\bibinfo {year} {2020})}\BibitemShut {NoStop}%
\bibitem [{\citenamefont {Ac{\'{\i}}n}\ \emph {et~al.}(2018)\citenamefont {Ac{\'{\i}}n}, \citenamefont {Bloch}, \citenamefont {Buhrman}, \citenamefont {Calarco}, \citenamefont {Eichler}, \citenamefont {Eisert}, \citenamefont {Esteve}, \citenamefont {Gisin}, \citenamefont {Glaser}, \citenamefont {Jelezko}, \citenamefont {Kuhr}, \citenamefont {Lewenstein}, \citenamefont {Riedel}, \citenamefont {Schmidt}, \citenamefont {Thew}, \citenamefont {Wallraff}, \citenamefont {Walmsley},\ and\ \citenamefont {Wilhelm}}]{Acin_2018}%
  \BibitemOpen
  \bibfield  {author} {\bibinfo {author} {\bibfnamefont {A.}~\bibnamefont {Ac{\'{\i}}n}}, \bibinfo {author} {\bibfnamefont {I.}~\bibnamefont {Bloch}}, \bibinfo {author} {\bibfnamefont {H.}~\bibnamefont {Buhrman}}, \bibinfo {author} {\bibfnamefont {T.}~\bibnamefont {Calarco}}, \bibinfo {author} {\bibfnamefont {C.}~\bibnamefont {Eichler}}, \bibinfo {author} {\bibfnamefont {J.}~\bibnamefont {Eisert}}, \bibinfo {author} {\bibfnamefont {D.}~\bibnamefont {Esteve}}, \bibinfo {author} {\bibfnamefont {N.}~\bibnamefont {Gisin}}, \bibinfo {author} {\bibfnamefont {S.~J.}\ \bibnamefont {Glaser}}, \bibinfo {author} {\bibfnamefont {F.}~\bibnamefont {Jelezko}}, \bibinfo {author} {\bibfnamefont {S.}~\bibnamefont {Kuhr}}, \bibinfo {author} {\bibfnamefont {M.}~\bibnamefont {Lewenstein}}, \bibinfo {author} {\bibfnamefont {M.~F.}\ \bibnamefont {Riedel}}, \bibinfo {author} {\bibfnamefont {P.~O.}\ \bibnamefont {Schmidt}}, \bibinfo {author} {\bibfnamefont {R.}~\bibnamefont {Thew}}, \bibinfo {author} {\bibfnamefont {A.}~\bibnamefont
  {Wallraff}}, \bibinfo {author} {\bibfnamefont {I.}~\bibnamefont {Walmsley}},\ and\ \bibinfo {author} {\bibfnamefont {F.~K.}\ \bibnamefont {Wilhelm}},\ }\bibfield  {title} {\bibinfo {title} {The quantum technologies roadmap: a european community view},\ }\href@noop {} {\bibfield  {journal} {\bibinfo  {journal} {New J. Phys.}\ }\textbf {\bibinfo {volume} {20}},\ \bibinfo {pages} {080201} (\bibinfo {year} {2018})}\BibitemShut {NoStop}%
\bibitem [{\citenamefont {Degen}\ \emph {et~al.}(2017)\citenamefont {Degen}, \citenamefont {Reinhard},\ and\ \citenamefont {Cappellaro}}]{Cappellaro_2017}%
  \BibitemOpen
  \bibfield  {author} {\bibinfo {author} {\bibfnamefont {C.~L.}\ \bibnamefont {Degen}}, \bibinfo {author} {\bibfnamefont {F.}~\bibnamefont {Reinhard}},\ and\ \bibinfo {author} {\bibfnamefont {P.}~\bibnamefont {Cappellaro}},\ }\bibfield  {title} {\bibinfo {title} {Quantum sensing},\ }\href {https://doi.org/10.1103/RevModPhys.89.035002} {\bibfield  {journal} {\bibinfo  {journal} {Rev. Mod. Phys.}\ }\textbf {\bibinfo {volume} {89}},\ \bibinfo {pages} {035002} (\bibinfo {year} {2017})}\BibitemShut {NoStop}%
\bibitem [{\citenamefont {Shapiro}\ and\ \citenamefont {Brumer}(2012)}]{Brumer2012}%
  \BibitemOpen
  \bibfield  {author} {\bibinfo {author} {\bibfnamefont {M.}~\bibnamefont {Shapiro}}\ and\ \bibinfo {author} {\bibfnamefont {P.}~\bibnamefont {Brumer}},\ }\href@noop {} {\emph {\bibinfo {title} {Quantum Control of Molecular Processes}}}\ (\bibinfo  {publisher} {(Wiley-VCH},\ \bibinfo {address} {Hoboken, NJ},\ \bibinfo {year} {2012})\BibitemShut {NoStop}%
\bibitem [{\citenamefont {Rice}\ and\ \citenamefont {Zhao}(2000)}]{ricebook}%
  \BibitemOpen
  \bibfield  {author} {\bibinfo {author} {\bibfnamefont {S.}~\bibnamefont {Rice}}\ and\ \bibinfo {author} {\bibfnamefont {M.}~\bibnamefont {Zhao}},\ }\href@noop {} {\emph {\bibinfo {title} {Optical Control of Molecular Dynamics}}}\ (\bibinfo  {publisher} {Wiley},\ \bibinfo {year} {2000})\BibitemShut {NoStop}%
\bibitem [{\citenamefont {Bayliss}\ \emph {et~al.}(2022)\citenamefont {Bayliss}, \citenamefont {Deb}, \citenamefont {Laorenza}, \citenamefont {Onizhuk}, \citenamefont {Galli}, \citenamefont {Freedman},\ and\ \citenamefont {Awschalom}}]{freedman2022}%
  \BibitemOpen
  \bibfield  {author} {\bibinfo {author} {\bibfnamefont {S.~L.}\ \bibnamefont {Bayliss}}, \bibinfo {author} {\bibfnamefont {P.}~\bibnamefont {Deb}}, \bibinfo {author} {\bibfnamefont {D.~W.}\ \bibnamefont {Laorenza}}, \bibinfo {author} {\bibfnamefont {M.}~\bibnamefont {Onizhuk}}, \bibinfo {author} {\bibfnamefont {G.}~\bibnamefont {Galli}}, \bibinfo {author} {\bibfnamefont {D.~E.}\ \bibnamefont {Freedman}},\ and\ \bibinfo {author} {\bibfnamefont {D.~D.}\ \bibnamefont {Awschalom}},\ }\bibfield  {title} {\bibinfo {title} {Enhancing spin coherence in optically addressable molecular qubits through host-matrix control},\ }\href {https://doi.org/10.1103/PhysRevX.12.031028} {\bibfield  {journal} {\bibinfo  {journal} {Phys. Rev. X}\ }\textbf {\bibinfo {volume} {12}},\ \bibinfo {pages} {031028} (\bibinfo {year} {2022})}\BibitemShut {NoStop}%
\bibitem [{\citenamefont {Schlosshauer}(2007)}]{Schlosshauer_2007}%
  \BibitemOpen
  \bibfield  {author} {\bibinfo {author} {\bibfnamefont {M.}~\bibnamefont {Schlosshauer}},\ }\href {https://books.google.com/books?id=1qrJUS5zNbEC} {\emph {\bibinfo {title} {Decoherence And the Quantum-To-Classical Transition}}},\ The Frontiers Collection\ (\bibinfo  {publisher} {Springer},\ \bibinfo {year} {2007})\BibitemShut {NoStop}%
\bibitem [{\citenamefont {Breuer}\ and\ \citenamefont {Petruccione}(2002)}]{Breuer2002}%
  \BibitemOpen
  \bibfield  {author} {\bibinfo {author} {\bibfnamefont {H.}~\bibnamefont {Breuer}}\ and\ \bibinfo {author} {\bibfnamefont {F.}~\bibnamefont {Petruccione}},\ }\href@noop {} {\emph {\bibinfo {title} {The Theory of Open Quantum Systems}}}\ (\bibinfo  {publisher} {Oxford University Press},\ \bibinfo {year} {2002})\BibitemShut {NoStop}%
\bibitem [{\citenamefont {Zhu}\ \emph {et~al.}(2022)\citenamefont {Zhu}, \citenamefont {Mitra}, \citenamefont {Augenbraun}, \citenamefont {Dickerson}, \citenamefont {Frim}, \citenamefont {Lao}, \citenamefont {Lasner}, \citenamefont {Alexandrova}, \citenamefont {Campbell}, \citenamefont {Caram} \emph {et~al.}}]{Caram2022}%
  \BibitemOpen
  \bibfield  {author} {\bibinfo {author} {\bibfnamefont {G.-Z.}\ \bibnamefont {Zhu}}, \bibinfo {author} {\bibfnamefont {D.}~\bibnamefont {Mitra}}, \bibinfo {author} {\bibfnamefont {B.~L.}\ \bibnamefont {Augenbraun}}, \bibinfo {author} {\bibfnamefont {C.~E.}\ \bibnamefont {Dickerson}}, \bibinfo {author} {\bibfnamefont {M.~J.}\ \bibnamefont {Frim}}, \bibinfo {author} {\bibfnamefont {G.}~\bibnamefont {Lao}}, \bibinfo {author} {\bibfnamefont {Z.~D.}\ \bibnamefont {Lasner}}, \bibinfo {author} {\bibfnamefont {A.~N.}\ \bibnamefont {Alexandrova}}, \bibinfo {author} {\bibfnamefont {W.~C.}\ \bibnamefont {Campbell}}, \bibinfo {author} {\bibfnamefont {J.~R.}\ \bibnamefont {Caram}}, \emph {et~al.},\ }\bibfield  {title} {\bibinfo {title} {Functionalizing aromatic compounds with optical cycling centres},\ }\href@noop {} {\bibfield  {journal} {\bibinfo  {journal} {Nat. Chem.}\ }\textbf {\bibinfo {volume} {14}},\ \bibinfo {pages} {995} (\bibinfo {year} {2022})}\BibitemShut {NoStop}%
\bibitem [{\citenamefont {Viola}\ \emph {et~al.}(1999)\citenamefont {Viola}, \citenamefont {Knill},\ and\ \citenamefont {Lloyd}}]{viola1999}%
  \BibitemOpen
  \bibfield  {author} {\bibinfo {author} {\bibfnamefont {L.}~\bibnamefont {Viola}}, \bibinfo {author} {\bibfnamefont {E.}~\bibnamefont {Knill}},\ and\ \bibinfo {author} {\bibfnamefont {S.}~\bibnamefont {Lloyd}},\ }\bibfield  {title} {\bibinfo {title} {Dynamical decoupling of open quantum systems},\ }\href@noop {} {\bibfield  {journal} {\bibinfo  {journal} {Phys. Rev. Lett.}\ }\textbf {\bibinfo {volume} {82}},\ \bibinfo {pages} {2417} (\bibinfo {year} {1999})}\BibitemShut {NoStop}%
\bibitem [{\citenamefont {Gustin}\ \emph {et~al.}(2023)\citenamefont {Gustin}, \citenamefont {Kim}, \citenamefont {McCamant},\ and\ \citenamefont {Franco}}]{Gustin2023}%
  \BibitemOpen
  \bibfield  {author} {\bibinfo {author} {\bibfnamefont {I.}~\bibnamefont {Gustin}}, \bibinfo {author} {\bibfnamefont {C.~W.}\ \bibnamefont {Kim}}, \bibinfo {author} {\bibfnamefont {D.~W.}\ \bibnamefont {McCamant}},\ and\ \bibinfo {author} {\bibfnamefont {I.}~\bibnamefont {Franco}},\ }\bibfield  {title} {\bibinfo {title} {Mapping electronic decoherence pathways in molecules},\ }\bibfield  {journal} {\bibinfo  {journal} {Proc. Natl. Acad. Sci.}\ }\textbf {\bibinfo {volume} {120}},\ \href {https://doi.org/10.1073/pnas.2309987120} {10.1073/pnas.2309987120} (\bibinfo {year} {2023})\BibitemShut {NoStop}%
\bibitem [{\citenamefont {Hwang}\ and\ \citenamefont {Rossky}(2004)}]{Hwang2004}%
  \BibitemOpen
  \bibfield  {author} {\bibinfo {author} {\bibfnamefont {H.}~\bibnamefont {Hwang}}\ and\ \bibinfo {author} {\bibfnamefont {P.~J.}\ \bibnamefont {Rossky}},\ }\bibfield  {title} {\bibinfo {title} {Electronic decoherence induced by intramolecular vibrational motions in a betaine dye molecule},\ }\href@noop {} {\bibfield  {journal} {\bibinfo  {journal} {J. Phys. Chem. B}\ }\textbf {\bibinfo {volume} {108}},\ \bibinfo {pages} {6723} (\bibinfo {year} {2004})}\BibitemShut {NoStop}%
\bibitem [{\citenamefont {Fleming}\ and\ \citenamefont {Wolynes}(1990)}]{Fleming_1990}%
  \BibitemOpen
  \bibfield  {author} {\bibinfo {author} {\bibfnamefont {G.~R.}\ \bibnamefont {Fleming}}\ and\ \bibinfo {author} {\bibfnamefont {P.~G.}\ \bibnamefont {Wolynes}},\ }\bibfield  {title} {\bibinfo {title} {Chemical dynamics in solution},\ }\href {https://doi.org/10.1063/1.881234} {\bibfield  {journal} {\bibinfo  {journal} {Phys. Today}\ }\textbf {\bibinfo {volume} {43}},\ \bibinfo {pages} {36} (\bibinfo {year} {1990})}\BibitemShut {NoStop}%
\bibitem [{\citenamefont {Zadrozny}\ \emph {et~al.}(2015)\citenamefont {Zadrozny}, \citenamefont {Niklas}, \citenamefont {Poluektov},\ and\ \citenamefont {Freedman}}]{Freedman_2015}%
  \BibitemOpen
  \bibfield  {author} {\bibinfo {author} {\bibfnamefont {J.~M.}\ \bibnamefont {Zadrozny}}, \bibinfo {author} {\bibfnamefont {J.}~\bibnamefont {Niklas}}, \bibinfo {author} {\bibfnamefont {O.~G.}\ \bibnamefont {Poluektov}},\ and\ \bibinfo {author} {\bibfnamefont {D.~E.}\ \bibnamefont {Freedman}},\ }\bibfield  {title} {\bibinfo {title} {Millisecond coherence time in a tunable molecular electronic spin qubit},\ }\href {https://doi.org/10.1021/acscentsci.5b00338} {\bibfield  {journal} {\bibinfo  {journal} {ACS Cent. Sci.}\ }\textbf {\bibinfo {volume} {1}},\ \bibinfo {pages} {488} (\bibinfo {year} {2015})}\BibitemShut {NoStop}%
\bibitem [{\citenamefont {Yu}\ \emph {et~al.}(2021)\citenamefont {Yu}, \citenamefont {von Kugelgen}, \citenamefont {Laorenza},\ and\ \citenamefont {Freedman}}]{Freedman_2021}%
  \BibitemOpen
  \bibfield  {author} {\bibinfo {author} {\bibfnamefont {C.-J.}\ \bibnamefont {Yu}}, \bibinfo {author} {\bibfnamefont {S.}~\bibnamefont {von Kugelgen}}, \bibinfo {author} {\bibfnamefont {D.~W.}\ \bibnamefont {Laorenza}},\ and\ \bibinfo {author} {\bibfnamefont {D.~E.}\ \bibnamefont {Freedman}},\ }\bibfield  {title} {\bibinfo {title} {A molecular approach to quantum sensing},\ }\href {https://doi.org/10.1021/acscentsci.0c00737} {\bibfield  {journal} {\bibinfo  {journal} {ACS Cent. Sci.}\ }\textbf {\bibinfo {volume} {7}},\ \bibinfo {pages} {712} (\bibinfo {year} {2021})}\BibitemShut {NoStop}%
\bibitem [{\citenamefont {Wang}\ \emph {et~al.}(2019{\natexlab{a}})\citenamefont {Wang}, \citenamefont {Kelkar}, \citenamefont {Martin-Cano}, \citenamefont {Rattenbacher}, \citenamefont {Shkarin}, \citenamefont {Utikal}, \citenamefont {G{\"o}tzinger},\ and\ \citenamefont {Sandoghdar}}]{wang2019turning}%
  \BibitemOpen
  \bibfield  {author} {\bibinfo {author} {\bibfnamefont {D.}~\bibnamefont {Wang}}, \bibinfo {author} {\bibfnamefont {H.}~\bibnamefont {Kelkar}}, \bibinfo {author} {\bibfnamefont {D.}~\bibnamefont {Martin-Cano}}, \bibinfo {author} {\bibfnamefont {D.}~\bibnamefont {Rattenbacher}}, \bibinfo {author} {\bibfnamefont {A.}~\bibnamefont {Shkarin}}, \bibinfo {author} {\bibfnamefont {T.}~\bibnamefont {Utikal}}, \bibinfo {author} {\bibfnamefont {S.}~\bibnamefont {G{\"o}tzinger}},\ and\ \bibinfo {author} {\bibfnamefont {V.}~\bibnamefont {Sandoghdar}},\ }\bibfield  {title} {\bibinfo {title} {Turning a molecule into a coherent two-level quantum system},\ }\href@noop {} {\bibfield  {journal} {\bibinfo  {journal} {Nat. Phys.}\ }\textbf {\bibinfo {volume} {15}},\ \bibinfo {pages} {483} (\bibinfo {year} {2019}{\natexlab{a}})}\BibitemShut {NoStop}%
\bibitem [{\citenamefont {Dickerson}\ \emph {et~al.}(2021)\citenamefont {Dickerson}, \citenamefont {Guo}, \citenamefont {Shin}, \citenamefont {Augenbraun}, \citenamefont {Caram}, \citenamefont {Campbell},\ and\ \citenamefont {Alexandrova}}]{dickerson2021}%
  \BibitemOpen
  \bibfield  {author} {\bibinfo {author} {\bibfnamefont {C.~E.}\ \bibnamefont {Dickerson}}, \bibinfo {author} {\bibfnamefont {H.}~\bibnamefont {Guo}}, \bibinfo {author} {\bibfnamefont {A.~J.}\ \bibnamefont {Shin}}, \bibinfo {author} {\bibfnamefont {B.~L.}\ \bibnamefont {Augenbraun}}, \bibinfo {author} {\bibfnamefont {J.~R.}\ \bibnamefont {Caram}}, \bibinfo {author} {\bibfnamefont {W.~C.}\ \bibnamefont {Campbell}},\ and\ \bibinfo {author} {\bibfnamefont {A.~N.}\ \bibnamefont {Alexandrova}},\ }\bibfield  {title} {\bibinfo {title} {Franck-condon tuning of optical cycling centers by organic functionalization},\ }\href@noop {} {\bibfield  {journal} {\bibinfo  {journal} {Phys. Rev. Lett.}\ }\textbf {\bibinfo {volume} {126}},\ \bibinfo {pages} {123002} (\bibinfo {year} {2021})}\BibitemShut {NoStop}%
\bibitem [{\citenamefont {Mukamel}(1995)}]{Mukamel_book}%
  \BibitemOpen
  \bibfield  {author} {\bibinfo {author} {\bibfnamefont {S.}~\bibnamefont {Mukamel}},\ }\href {https://books.google.com/books?id=k_7uAAAAMAAJ} {\emph {\bibinfo {title} {Principles of Nonlinear Optical Spectroscopy}}}\ (\bibinfo  {publisher} {Oxford University Press},\ \bibinfo {year} {1995})\BibitemShut {NoStop}%
\bibitem [{\citenamefont {Wang}\ \emph {et~al.}(2019{\natexlab{b}})\citenamefont {Wang}, \citenamefont {Allodi},\ and\ \citenamefont {Engel}}]{Engel_2019}%
  \BibitemOpen
  \bibfield  {author} {\bibinfo {author} {\bibfnamefont {L.}~\bibnamefont {Wang}}, \bibinfo {author} {\bibfnamefont {M.~A.}\ \bibnamefont {Allodi}},\ and\ \bibinfo {author} {\bibfnamefont {G.~S.}\ \bibnamefont {Engel}},\ }\bibfield  {title} {\bibinfo {title} {Quantum coherences reveal excited-state dynamics in biophysical systems},\ }\href@noop {} {\bibfield  {journal} {\bibinfo  {journal} {Nat. Rev. Chem.}\ }\textbf {\bibinfo {volume} {3}},\ \bibinfo {pages} {477} (\bibinfo {year} {2019}{\natexlab{b}})}\BibitemShut {NoStop}%
\bibitem [{\citenamefont {Gururangan}\ and\ \citenamefont {Harel}(2019)}]{Harel_2019}%
  \BibitemOpen
  \bibfield  {author} {\bibinfo {author} {\bibfnamefont {K.}~\bibnamefont {Gururangan}}\ and\ \bibinfo {author} {\bibfnamefont {E.}~\bibnamefont {Harel}},\ }\bibfield  {title} {\bibinfo {title} {Coherent and dissipative quantum process tensor reconstructions in two-dimensional electronic spectroscopy},\ }\href {https://doi.org/10.1063/1.5082165} {\bibfield  {journal} {\bibinfo  {journal} {J. Chem. Phys.}\ }\textbf {\bibinfo {volume} {150}},\ \bibinfo {pages} {164127} (\bibinfo {year} {2019})}\BibitemShut {NoStop}%
\bibitem [{\citenamefont {Unruh}(1995)}]{unruh1995}%
  \BibitemOpen
  \bibfield  {author} {\bibinfo {author} {\bibfnamefont {W.~G.}\ \bibnamefont {Unruh}},\ }\bibfield  {title} {\bibinfo {title} {Maintaining coherence in quantum computers},\ }\href@noop {} {\bibfield  {journal} {\bibinfo  {journal} {Phys. Rev. A}\ }\textbf {\bibinfo {volume} {51}},\ \bibinfo {pages} {992} (\bibinfo {year} {1995})}\BibitemShut {NoStop}%
\bibitem [{\citenamefont {Palma}\ \emph {et~al.}(1996)\citenamefont {Palma}, \citenamefont {Suominen},\ and\ \citenamefont {Ekert}}]{palma1996}%
  \BibitemOpen
  \bibfield  {author} {\bibinfo {author} {\bibfnamefont {G.~M.}\ \bibnamefont {Palma}}, \bibinfo {author} {\bibfnamefont {K.-A.}\ \bibnamefont {Suominen}},\ and\ \bibinfo {author} {\bibfnamefont {A.}~\bibnamefont {Ekert}},\ }\bibfield  {title} {\bibinfo {title} {Quantum computers and dissipation},\ }\href {https://doi.org/10.1098/rspa.1996.0029} {\bibfield  {journal} {\bibinfo  {journal} {Proc. R. Soc. Lond.}\ }\textbf {\bibinfo {volume} {A452}},\ \bibinfo {pages} {567} (\bibinfo {year} {1996})}\BibitemShut {NoStop}%
\bibitem [{\citenamefont {Brinks}\ \emph {et~al.}(2014)\citenamefont {Brinks}, \citenamefont {Hildner}, \citenamefont {Van~Dijk}, \citenamefont {Stefani}, \citenamefont {Nieder}, \citenamefont {Hernando},\ and\ \citenamefont {Van~Hulst}}]{brinks2014ultrafast}%
  \BibitemOpen
  \bibfield  {author} {\bibinfo {author} {\bibfnamefont {D.}~\bibnamefont {Brinks}}, \bibinfo {author} {\bibfnamefont {R.}~\bibnamefont {Hildner}}, \bibinfo {author} {\bibfnamefont {E.~M.}\ \bibnamefont {Van~Dijk}}, \bibinfo {author} {\bibfnamefont {F.~D.}\ \bibnamefont {Stefani}}, \bibinfo {author} {\bibfnamefont {J.~B.}\ \bibnamefont {Nieder}}, \bibinfo {author} {\bibfnamefont {J.}~\bibnamefont {Hernando}},\ and\ \bibinfo {author} {\bibfnamefont {N.~F.}\ \bibnamefont {Van~Hulst}},\ }\bibfield  {title} {\bibinfo {title} {Ultrafast dynamics of single molecules},\ }\href@noop {} {\bibfield  {journal} {\bibinfo  {journal} {Chem. Soc. Rev.}\ }\textbf {\bibinfo {volume} {43}},\ \bibinfo {pages} {2476} (\bibinfo {year} {2014})}\BibitemShut {NoStop}%
\bibitem [{\citenamefont {Von~der Linde}\ \emph {et~al.}(1997)\citenamefont {Von~der Linde}, \citenamefont {Sokolowski-Tinten},\ and\ \citenamefont {Bialkowski}}]{von1997laser}%
  \BibitemOpen
  \bibfield  {author} {\bibinfo {author} {\bibfnamefont {D.}~\bibnamefont {Von~der Linde}}, \bibinfo {author} {\bibfnamefont {K.}~\bibnamefont {Sokolowski-Tinten}},\ and\ \bibinfo {author} {\bibfnamefont {J.}~\bibnamefont {Bialkowski}},\ }\bibfield  {title} {\bibinfo {title} {Laser--solid interaction in the femtosecond time regime},\ }\href@noop {} {\bibfield  {journal} {\bibinfo  {journal} {Appl. Surf. Sci.}\ }\textbf {\bibinfo {volume} {109}},\ \bibinfo {pages} {1} (\bibinfo {year} {1997})}\BibitemShut {NoStop}%
\bibitem [{\citenamefont {Feynman}\ and\ \citenamefont {Vernon}(1963)}]{Feynman1963}%
  \BibitemOpen
  \bibfield  {author} {\bibinfo {author} {\bibfnamefont {R.~P.}\ \bibnamefont {Feynman}}\ and\ \bibinfo {author} {\bibfnamefont {F.~L.}\ \bibnamefont {Vernon}},\ }\bibfield  {title} {\bibinfo {title} {The theory of a general quantum system interacting with a linear dissipative system},\ }\href {https://doi.org/10.1016/0003-4916(63)90068-X} {\bibfield  {journal} {\bibinfo  {journal} {Ann. Phys.}\ }\textbf {\bibinfo {volume} {24}},\ \bibinfo {pages} {118} (\bibinfo {year} {1963})}\BibitemShut {NoStop}%
\bibitem [{\citenamefont {Caldeira}\ and\ \citenamefont {Leggett}(1983)}]{Caldeira1983}%
  \BibitemOpen
  \bibfield  {author} {\bibinfo {author} {\bibfnamefont {A.~O.}\ \bibnamefont {Caldeira}}\ and\ \bibinfo {author} {\bibfnamefont {A.~J.}\ \bibnamefont {Leggett}},\ }\bibfield  {title} {\bibinfo {title} {Quantum tunnelling in a dissipative system},\ }\href {https://doi.org/10.1016/0003-4916(83)90202-6} {\bibfield  {journal} {\bibinfo  {journal} {Ann. Phys.}\ }\textbf {\bibinfo {volume} {149}},\ \bibinfo {pages} {374} (\bibinfo {year} {1983})}\BibitemShut {NoStop}%
\bibitem [{\citenamefont {Leggett}\ \emph {et~al.}(1987)\citenamefont {Leggett}, \citenamefont {Chakravarty}, \citenamefont {Dorsey}, \citenamefont {Fisher}, \citenamefont {Garg},\ and\ \citenamefont {Zwerger}}]{Leggett1987}%
  \BibitemOpen
  \bibfield  {author} {\bibinfo {author} {\bibfnamefont {A.~J.}\ \bibnamefont {Leggett}}, \bibinfo {author} {\bibfnamefont {S.}~\bibnamefont {Chakravarty}}, \bibinfo {author} {\bibfnamefont {A.~T.}\ \bibnamefont {Dorsey}}, \bibinfo {author} {\bibfnamefont {M.~P.~A.}\ \bibnamefont {Fisher}}, \bibinfo {author} {\bibfnamefont {A.}~\bibnamefont {Garg}},\ and\ \bibinfo {author} {\bibfnamefont {W.}~\bibnamefont {Zwerger}},\ }\bibfield  {title} {\bibinfo {title} {Dynamics of the dissipative two-state system},\ }\href {https://doi.org/10.1103/RevModPhys.59.1} {\bibfield  {journal} {\bibinfo  {journal} {Rev. Mod. Phys.}\ }\textbf {\bibinfo {volume} {59}},\ \bibinfo {pages} {1} (\bibinfo {year} {1987})}\BibitemShut {NoStop}%
\bibitem [{\citenamefont {Caldeira}\ \emph {et~al.}(1993)\citenamefont {Caldeira}, \citenamefont {CastroNeto},\ and\ \citenamefont {de~Carvalho}}]{Caldeira1993}%
  \BibitemOpen
  \bibfield  {author} {\bibinfo {author} {\bibfnamefont {A.~O.}\ \bibnamefont {Caldeira}}, \bibinfo {author} {\bibfnamefont {A.~H.}\ \bibnamefont {CastroNeto}},\ and\ \bibinfo {author} {\bibfnamefont {T.~O.}\ \bibnamefont {de~Carvalho}},\ }\bibfield  {title} {\bibinfo {title} {Dissipative quantum systems modeled by a two-level-reservoir coupling},\ }\href {https://doi.org/10.1103/PhysRevB.48.13974} {\bibfield  {journal} {\bibinfo  {journal} {Phys. Rev. B}\ }\textbf {\bibinfo {volume} {48}},\ \bibinfo {pages} {13974} (\bibinfo {year} {1993})}\BibitemShut {NoStop}%
\bibitem [{\citenamefont {Wiethorn}\ \emph {et~al.}(2023)\citenamefont {Wiethorn}, \citenamefont {Hunter}, \citenamefont {Zuehlsdorff},\ and\ \citenamefont {Montoya-Castillo}}]{wiethorn2023beyond}%
  \BibitemOpen
  \bibfield  {author} {\bibinfo {author} {\bibfnamefont {Z.~R.}\ \bibnamefont {Wiethorn}}, \bibinfo {author} {\bibfnamefont {K.~E.}\ \bibnamefont {Hunter}}, \bibinfo {author} {\bibfnamefont {T.~J.}\ \bibnamefont {Zuehlsdorff}},\ and\ \bibinfo {author} {\bibfnamefont {A.}~\bibnamefont {Montoya-Castillo}},\ }\bibfield  {title} {\bibinfo {title} {Beyond the condon limit: Condensed phase optical spectra from atomistic simulations},\ }\href@noop {} {\bibfield  {journal} {\bibinfo  {journal} {J. Chem. Phys.}\ }\textbf {\bibinfo {volume} {159}} (\bibinfo {year} {2023})}\BibitemShut {NoStop}%
\bibitem [{\citenamefont {Lunghi}\ and\ \citenamefont {Sanvito}(2019)}]{lunghi2019phonons}%
  \BibitemOpen
  \bibfield  {author} {\bibinfo {author} {\bibfnamefont {A.}~\bibnamefont {Lunghi}}\ and\ \bibinfo {author} {\bibfnamefont {S.}~\bibnamefont {Sanvito}},\ }\bibfield  {title} {\bibinfo {title} {How do phonons relax molecular spins?},\ }\href@noop {} {\bibfield  {journal} {\bibinfo  {journal} {Sci. adv.}\ }\textbf {\bibinfo {volume} {5}},\ \bibinfo {pages} {eaax7163} (\bibinfo {year} {2019})}\BibitemShut {NoStop}%
\bibitem [{\citenamefont {Garlatti}\ \emph {et~al.}(2023)\citenamefont {Garlatti}, \citenamefont {Albino}, \citenamefont {Chicco}, \citenamefont {Nguyen}, \citenamefont {Santanni}, \citenamefont {Paolasini}, \citenamefont {Mazzoli}, \citenamefont {Caciuffo}, \citenamefont {Totti}, \citenamefont {Santini} \emph {et~al.}}]{garlatti2023critical}%
  \BibitemOpen
  \bibfield  {author} {\bibinfo {author} {\bibfnamefont {E.}~\bibnamefont {Garlatti}}, \bibinfo {author} {\bibfnamefont {A.}~\bibnamefont {Albino}}, \bibinfo {author} {\bibfnamefont {S.}~\bibnamefont {Chicco}}, \bibinfo {author} {\bibfnamefont {V.}~\bibnamefont {Nguyen}}, \bibinfo {author} {\bibfnamefont {F.}~\bibnamefont {Santanni}}, \bibinfo {author} {\bibfnamefont {L.}~\bibnamefont {Paolasini}}, \bibinfo {author} {\bibfnamefont {C.}~\bibnamefont {Mazzoli}}, \bibinfo {author} {\bibfnamefont {R.}~\bibnamefont {Caciuffo}}, \bibinfo {author} {\bibfnamefont {F.}~\bibnamefont {Totti}}, \bibinfo {author} {\bibfnamefont {P.}~\bibnamefont {Santini}}, \emph {et~al.},\ }\bibfield  {title} {\bibinfo {title} {The critical role of ultra-low-energy vibrations in the relaxation dynamics of molecular qubits},\ }\href@noop {} {\bibfield  {journal} {\bibinfo  {journal} {Nat. Commun.}\ }\textbf {\bibinfo {volume} {14}},\ \bibinfo {pages} {1653} (\bibinfo {year} {2023})}\BibitemShut {NoStop}%
\bibitem [{\citenamefont {Lunghi}(2022)}]{lunghi2022toward}%
  \BibitemOpen
  \bibfield  {author} {\bibinfo {author} {\bibfnamefont {A.}~\bibnamefont {Lunghi}},\ }\bibfield  {title} {\bibinfo {title} {Toward exact predictions of spin-phonon relaxation times: An ab initio implementation of open quantum systems theory},\ }\href@noop {} {\bibfield  {journal} {\bibinfo  {journal} {Sci. Adv}\ }\textbf {\bibinfo {volume} {8}},\ \bibinfo {pages} {eabn7880} (\bibinfo {year} {2022})}\BibitemShut {NoStop}%
\bibitem [{\citenamefont {Vezvaee}\ \emph {et~al.}(2024)\citenamefont {Vezvaee}, \citenamefont {Shitara}, \citenamefont {Sun},\ and\ \citenamefont {Montoya-Castillo}}]{Vezvaee2024}%
  \BibitemOpen
  \bibfield  {author} {\bibinfo {author} {\bibfnamefont {A.}~\bibnamefont {Vezvaee}}, \bibinfo {author} {\bibfnamefont {N.}~\bibnamefont {Shitara}}, \bibinfo {author} {\bibfnamefont {S.}~\bibnamefont {Sun}},\ and\ \bibinfo {author} {\bibfnamefont {A.}~\bibnamefont {Montoya-Castillo}},\ }\bibfield  {title} {\bibinfo {title} {Fourier transform noise spectroscopy},\ }\href {https://doi.org/10.1038/s41534-024-00841-w} {\bibfield  {journal} {\bibinfo  {journal} {npj Quantum Information}\ }\textbf {\bibinfo {volume} {10}},\ \bibinfo {pages} {52} (\bibinfo {year} {2024})}\BibitemShut {NoStop}%
\bibitem [{\citenamefont {Paz-Silva}\ \emph {et~al.}(2017)\citenamefont {Paz-Silva}, \citenamefont {Norris},\ and\ \citenamefont {Viola}}]{paz2017multiqubit}%
  \BibitemOpen
  \bibfield  {author} {\bibinfo {author} {\bibfnamefont {G.~A.}\ \bibnamefont {Paz-Silva}}, \bibinfo {author} {\bibfnamefont {L.~M.}\ \bibnamefont {Norris}},\ and\ \bibinfo {author} {\bibfnamefont {L.}~\bibnamefont {Viola}},\ }\bibfield  {title} {\bibinfo {title} {Multiqubit spectroscopy of gaussian quantum noise},\ }\href@noop {} {\bibfield  {journal} {\bibinfo  {journal} {Phys. Rev. A}\ }\textbf {\bibinfo {volume} {95}},\ \bibinfo {pages} {022121} (\bibinfo {year} {2017})}\BibitemShut {NoStop}%
\bibitem [{\citenamefont {Kwiatkowski}\ \emph {et~al.}(2020)\citenamefont {Kwiatkowski}, \citenamefont {Sza{\'n}kowski},\ and\ \citenamefont {Cywi{\'n}ski}}]{kwiatkowski2020influence}%
  \BibitemOpen
  \bibfield  {author} {\bibinfo {author} {\bibfnamefont {D.}~\bibnamefont {Kwiatkowski}}, \bibinfo {author} {\bibfnamefont {P.}~\bibnamefont {Sza{\'n}kowski}},\ and\ \bibinfo {author} {\bibfnamefont {{\L}.}~\bibnamefont {Cywi{\'n}ski}},\ }\bibfield  {title} {\bibinfo {title} {Influence of nuclear spin polarization on the spin-echo signal of an nv-center qubit},\ }\href@noop {} {\bibfield  {journal} {\bibinfo  {journal} {Phys. Rev. B}\ }\textbf {\bibinfo {volume} {101}},\ \bibinfo {pages} {155412} (\bibinfo {year} {2020})}\BibitemShut {NoStop}%
\bibitem [{\citenamefont {Cywi{\'n}ski}\ \emph {et~al.}(2008)\citenamefont {Cywi{\'n}ski}, \citenamefont {Lutchyn}, \citenamefont {Nave},\ and\ \citenamefont {Das~Sarma}}]{cywinski2008enhance}%
  \BibitemOpen
  \bibfield  {author} {\bibinfo {author} {\bibfnamefont {{\L}.}~\bibnamefont {Cywi{\'n}ski}}, \bibinfo {author} {\bibfnamefont {R.~M.}\ \bibnamefont {Lutchyn}}, \bibinfo {author} {\bibfnamefont {C.~P.}\ \bibnamefont {Nave}},\ and\ \bibinfo {author} {\bibfnamefont {S.}~\bibnamefont {Das~Sarma}},\ }\bibfield  {title} {\bibinfo {title} {How to enhance dephasing time in superconducting qubits},\ }\href@noop {} {\bibfield  {journal} {\bibinfo  {journal} {Phys. Rev. B}\ }\textbf {\bibinfo {volume} {77}},\ \bibinfo {pages} {174509} (\bibinfo {year} {2008})}\BibitemShut {NoStop}%
\bibitem [{\citenamefont {Schatz}\ and\ \citenamefont {Ratner}(2002)}]{schatz2002quantum}%
  \BibitemOpen
  \bibfield  {author} {\bibinfo {author} {\bibfnamefont {G.~C.}\ \bibnamefont {Schatz}}\ and\ \bibinfo {author} {\bibfnamefont {M.~A.}\ \bibnamefont {Ratner}},\ }\href@noop {} {\emph {\bibinfo {title} {Quantum mechanics in chemistry}}}\ (\bibinfo  {publisher} {Courier Corporation},\ \bibinfo {year} {2002})\BibitemShut {NoStop}%
\bibitem [{\citenamefont {Skinner}\ and\ \citenamefont {Hsu}(1986)}]{skinner1986pure}%
  \BibitemOpen
  \bibfield  {author} {\bibinfo {author} {\bibfnamefont {J.}~\bibnamefont {Skinner}}\ and\ \bibinfo {author} {\bibfnamefont {D.}~\bibnamefont {Hsu}},\ }\bibfield  {title} {\bibinfo {title} {Pure dephasing of a two-level system},\ }\href@noop {} {\bibfield  {journal} {\bibinfo  {journal} {J. Phys. Chem.}\ }\textbf {\bibinfo {volume} {90}},\ \bibinfo {pages} {4931} (\bibinfo {year} {1986})}\BibitemShut {NoStop}%
\bibitem [{\citenamefont {Tokmakoff}(2014)}]{tokmakotime}%
  \BibitemOpen
  \bibfield  {author} {\bibinfo {author} {\bibfnamefont {A.}~\bibnamefont {Tokmakoff}},\ }\href {https://books.google.com/books?id=Vd1P0AEACAAJ} {\emph {\bibinfo {title} {Time Dependent Quantum Mechanics and Spectroscopy}}}\ (\bibinfo  {publisher} {LibreTexts},\ \bibinfo {year} {2014})\BibitemShut {NoStop}%
\bibitem [{\citenamefont {Hu}\ \emph {et~al.}(2018)\citenamefont {Hu}, \citenamefont {Gu},\ and\ \citenamefont {Franco}}]{Wen2018}%
  \BibitemOpen
  \bibfield  {author} {\bibinfo {author} {\bibfnamefont {W.}~\bibnamefont {Hu}}, \bibinfo {author} {\bibfnamefont {B.}~\bibnamefont {Gu}},\ and\ \bibinfo {author} {\bibfnamefont {I.}~\bibnamefont {Franco}},\ }\bibfield  {title} {\bibinfo {title} {Lessons on electronic decoherence in molecules from exact modeling},\ }\href {https://doi.org/10.1063/1.5004578} {\bibfield  {journal} {\bibinfo  {journal} {J. Chem. Phys.}\ }\textbf {\bibinfo {volume} {148}},\ \bibinfo {pages} {134304} (\bibinfo {year} {2018})}\BibitemShut {NoStop}%
\bibitem [{\citenamefont {Nitzan}(2006)}]{Nitzan2006}%
  \BibitemOpen
  \bibfield  {author} {\bibinfo {author} {\bibfnamefont {A.}~\bibnamefont {Nitzan}},\ }\href@noop {} {\emph {\bibinfo {title} {Chemical {{Dynamics}} in {{Condensed Phases}}: {{Relaxation}}, {{Transfer}} and {{Reactions}} in {{Condensed Molecular Systems}}}}}\ (\bibinfo  {publisher} {{Oxford University Press}},\ \bibinfo {year} {2006})\BibitemShut {NoStop}%
\bibitem [{\citenamefont {Gu}\ and\ \citenamefont {Franco}(2019)}]{Gu_2019}%
  \BibitemOpen
  \bibfield  {author} {\bibinfo {author} {\bibfnamefont {B.}~\bibnamefont {Gu}}\ and\ \bibinfo {author} {\bibfnamefont {I.}~\bibnamefont {Franco}},\ }\bibfield  {title} {\bibinfo {title} {When can quantum decoherence be mimicked by classical noise?},\ }\href {https://doi.org/10.1063/1.5099499} {\bibfield  {journal} {\bibinfo  {journal} {J. Chem. Phys.}\ }\textbf {\bibinfo {volume} {151}},\ \bibinfo {pages} {014109} (\bibinfo {year} {2019})}\BibitemShut {NoStop}%
\bibitem [{\citenamefont {Kubo}(1969)}]{kubo1969}%
  \BibitemOpen
  \bibfield  {author} {\bibinfo {author} {\bibfnamefont {R.}~\bibnamefont {Kubo}},\ }\bibfield  {title} {\bibinfo {title} {A stochastic theory of line shape},\ }\href@noop {} {\bibfield  {journal} {\bibinfo  {journal} {Adv. Chem. Phys.}\ }\textbf {\bibinfo {volume} {15}},\ \bibinfo {pages} {101} (\bibinfo {year} {1969})}\BibitemShut {NoStop}%
\bibitem [{\citenamefont {Joutsuka}\ \emph {et~al.}(2016)\citenamefont {Joutsuka}, \citenamefont {Thompson},\ and\ \citenamefont {Laage}}]{Joutsuka2016}%
  \BibitemOpen
  \bibfield  {author} {\bibinfo {author} {\bibfnamefont {T.}~\bibnamefont {Joutsuka}}, \bibinfo {author} {\bibfnamefont {W.~H.}\ \bibnamefont {Thompson}},\ and\ \bibinfo {author} {\bibfnamefont {D.}~\bibnamefont {Laage}},\ }\bibfield  {title} {\bibinfo {title} {Vibrational quantum decoherence in liquid water},\ }\href@noop {} {\bibfield  {journal} {\bibinfo  {journal} {J. Phys. Chem. Lett.}\ }\textbf {\bibinfo {volume} {7}},\ \bibinfo {pages} {616} (\bibinfo {year} {2016})}\BibitemShut {NoStop}%
\bibitem [{\citenamefont {Yang}\ \emph {et~al.}(2016)\citenamefont {Yang}, \citenamefont {Ma},\ and\ \citenamefont {Liu}}]{yang2016quantum}%
  \BibitemOpen
  \bibfield  {author} {\bibinfo {author} {\bibfnamefont {W.}~\bibnamefont {Yang}}, \bibinfo {author} {\bibfnamefont {W.-L.}\ \bibnamefont {Ma}},\ and\ \bibinfo {author} {\bibfnamefont {R.-B.}\ \bibnamefont {Liu}},\ }\bibfield  {title} {\bibinfo {title} {Quantum many-body theory for electron spin decoherence in nanoscale nuclear spin baths},\ }\href@noop {} {\bibfield  {journal} {\bibinfo  {journal} {Rep. Prog. Phys.}\ }\textbf {\bibinfo {volume} {80}},\ \bibinfo {pages} {016001} (\bibinfo {year} {2016})}\BibitemShut {NoStop}%
\bibitem [{\citenamefont {Mukamel}(1985)}]{mukamel1985fluorescence}%
  \BibitemOpen
  \bibfield  {author} {\bibinfo {author} {\bibfnamefont {S.}~\bibnamefont {Mukamel}},\ }\bibfield  {title} {\bibinfo {title} {Fluorescence and absorption of large anharmonic molecules-spectroscopy without eigenstates},\ }\href@noop {} {\bibfield  {journal} {\bibinfo  {journal} {J. Phys. Chem.}\ }\textbf {\bibinfo {volume} {89}},\ \bibinfo {pages} {1077} (\bibinfo {year} {1985})}\BibitemShut {NoStop}%
\bibitem [{\citenamefont {Cai}\ and\ \citenamefont {Barthel}(2013)}]{cai2013}%
  \BibitemOpen
  \bibfield  {author} {\bibinfo {author} {\bibfnamefont {Z.}~\bibnamefont {Cai}}\ and\ \bibinfo {author} {\bibfnamefont {T.}~\bibnamefont {Barthel}},\ }\bibfield  {title} {\bibinfo {title} {Algebraic versus exponential decoherence in dissipative many-particle systems},\ }\href@noop {} {\bibfield  {journal} {\bibinfo  {journal} {Phys. Rev. Lett.}\ }\textbf {\bibinfo {volume} {111}},\ \bibinfo {pages} {150403} (\bibinfo {year} {2013})}\BibitemShut {NoStop}%
\bibitem [{\citenamefont {Knight}(1976)}]{knight1976}%
  \BibitemOpen
  \bibfield  {author} {\bibinfo {author} {\bibfnamefont {P.}~\bibnamefont {Knight}},\ }\bibfield  {title} {\bibinfo {title} {Ersaks's regeneration hypothesis and deviations from the exponential decay law},\ }\href@noop {} {\bibfield  {journal} {\bibinfo  {journal} {Phys. Lett. A}\ }\textbf {\bibinfo {volume} {56}},\ \bibinfo {pages} {11} (\bibinfo {year} {1976})}\BibitemShut {NoStop}%
\bibitem [{\citenamefont {Burgarth}\ and\ \citenamefont {Facchi}(2017)}]{burgarth2017}%
  \BibitemOpen
  \bibfield  {author} {\bibinfo {author} {\bibfnamefont {D.}~\bibnamefont {Burgarth}}\ and\ \bibinfo {author} {\bibfnamefont {P.}~\bibnamefont {Facchi}},\ }\bibfield  {title} {\bibinfo {title} {Positive hamiltonians can give purely exponential decay},\ }\href@noop {} {\bibfield  {journal} {\bibinfo  {journal} {Phys. Rev. A}\ }\textbf {\bibinfo {volume} {96}},\ \bibinfo {pages} {010103} (\bibinfo {year} {2017})}\BibitemShut {NoStop}%
\bibitem [{\citenamefont {Unruh}\ and\ \citenamefont {Zurek}(1989)}]{unruh1989}%
  \BibitemOpen
  \bibfield  {author} {\bibinfo {author} {\bibfnamefont {W.}~\bibnamefont {Unruh}}\ and\ \bibinfo {author} {\bibfnamefont {W.~H.}\ \bibnamefont {Zurek}},\ }\bibfield  {title} {\bibinfo {title} {Reduction of a wave packet in quantum brownian motion},\ }\href@noop {} {\bibfield  {journal} {\bibinfo  {journal} {Phys. Rev. D}\ }\textbf {\bibinfo {volume} {40}},\ \bibinfo {pages} {1071} (\bibinfo {year} {1989})}\BibitemShut {NoStop}%
\bibitem [{\citenamefont {Paz}\ \emph {et~al.}(1993)\citenamefont {Paz}, \citenamefont {Habib},\ and\ \citenamefont {Zurek}}]{paz1993}%
  \BibitemOpen
  \bibfield  {author} {\bibinfo {author} {\bibfnamefont {J.~P.}\ \bibnamefont {Paz}}, \bibinfo {author} {\bibfnamefont {S.}~\bibnamefont {Habib}},\ and\ \bibinfo {author} {\bibfnamefont {W.~H.}\ \bibnamefont {Zurek}},\ }\bibfield  {title} {\bibinfo {title} {Reduction of the wave packet: Preferred observable and decoherence time scale},\ }\href@noop {} {\bibfield  {journal} {\bibinfo  {journal} {Phys. Rev. D}\ }\textbf {\bibinfo {volume} {47}},\ \bibinfo {pages} {488} (\bibinfo {year} {1993})}\BibitemShut {NoStop}%
\bibitem [{\citenamefont {Zurek}(2003)}]{zurek2003}%
  \BibitemOpen
  \bibfield  {author} {\bibinfo {author} {\bibfnamefont {W.~H.}\ \bibnamefont {Zurek}},\ }\bibfield  {title} {\bibinfo {title} {Decoherence, einselection, and the quantum origins of the classical},\ }\href@noop {} {\bibfield  {journal} {\bibinfo  {journal} {Rev. Mod. Phys.}\ }\textbf {\bibinfo {volume} {75}},\ \bibinfo {pages} {715} (\bibinfo {year} {2003})}\BibitemShut {NoStop}%
\bibitem [{\citenamefont {Xu}\ \emph {et~al.}(2019)\citenamefont {Xu}, \citenamefont {Garc{\'\i}a-Pintos}, \citenamefont {Chenu},\ and\ \citenamefont {Del~Campo}}]{xu2019}%
  \BibitemOpen
  \bibfield  {author} {\bibinfo {author} {\bibfnamefont {Z.}~\bibnamefont {Xu}}, \bibinfo {author} {\bibfnamefont {L.~P.}\ \bibnamefont {Garc{\'\i}a-Pintos}}, \bibinfo {author} {\bibfnamefont {A.}~\bibnamefont {Chenu}},\ and\ \bibinfo {author} {\bibfnamefont {A.}~\bibnamefont {Del~Campo}},\ }\bibfield  {title} {\bibinfo {title} {Extreme decoherence and quantum chaos},\ }\href@noop {} {\bibfield  {journal} {\bibinfo  {journal} {Phys. Rev. Lett.}\ }\textbf {\bibinfo {volume} {122}},\ \bibinfo {pages} {014103} (\bibinfo {year} {2019})}\BibitemShut {NoStop}%
\bibitem [{\citenamefont {Hsu}\ and\ \citenamefont {Skinner}(1984)}]{hsu1984thermal}%
  \BibitemOpen
  \bibfield  {author} {\bibinfo {author} {\bibfnamefont {D.}~\bibnamefont {Hsu}}\ and\ \bibinfo {author} {\bibfnamefont {J.}~\bibnamefont {Skinner}},\ }\bibfield  {title} {\bibinfo {title} {On the thermal broadening of zero-phonon impurity lines in absorption and fluorescence spectra},\ }\href@noop {} {\bibfield  {journal} {\bibinfo  {journal} {J. Chem. Phys.}\ }\textbf {\bibinfo {volume} {81}},\ \bibinfo {pages} {1604} (\bibinfo {year} {1984})}\BibitemShut {NoStop}%
\bibitem [{\citenamefont {Hansen}\ and\ \citenamefont {McDonald}(2013)}]{hansen2013theory}%
  \BibitemOpen
  \bibfield  {author} {\bibinfo {author} {\bibfnamefont {J.-P.}\ \bibnamefont {Hansen}}\ and\ \bibinfo {author} {\bibfnamefont {I.~R.}\ \bibnamefont {McDonald}},\ }\href@noop {} {\emph {\bibinfo {title} {Theory of simple liquids: with applications to soft matter}}}\ (\bibinfo  {publisher} {Academic press},\ \bibinfo {year} {2013})\BibitemShut {NoStop}%
\bibitem [{\citenamefont {Gu}\ and\ \citenamefont {Franco}(2018)}]{Bing2018}%
  \BibitemOpen
  \bibfield  {author} {\bibinfo {author} {\bibfnamefont {B.}~\bibnamefont {Gu}}\ and\ \bibinfo {author} {\bibfnamefont {I.}~\bibnamefont {Franco}},\ }\bibfield  {title} {\bibinfo {title} {Generalized theory for the timescale of molecular electronic decoherence in the condensed phase},\ }\href {https://doi.org/10.1021/acs.jpclett.7b03322} {\bibfield  {journal} {\bibinfo  {journal} {J. Phys. Chem. Lett.}\ }\textbf {\bibinfo {volume} {9}},\ \bibinfo {pages} {773} (\bibinfo {year} {2018})}\BibitemShut {NoStop}%
\bibitem [{\citenamefont {Hu}\ \emph {et~al.}(2022)\citenamefont {Hu}, \citenamefont {Gustin}, \citenamefont {Krauss},\ and\ \citenamefont {Franco}}]{Wen2022}%
  \BibitemOpen
  \bibfield  {author} {\bibinfo {author} {\bibfnamefont {W.}~\bibnamefont {Hu}}, \bibinfo {author} {\bibfnamefont {I.}~\bibnamefont {Gustin}}, \bibinfo {author} {\bibfnamefont {T.~D.}\ \bibnamefont {Krauss}},\ and\ \bibinfo {author} {\bibfnamefont {I.}~\bibnamefont {Franco}},\ }\bibfield  {title} {\bibinfo {title} {Tuning and enhancing quantum coherence time scales in molecules via light-matter hybridization},\ }\href@noop {} {\bibfield  {journal} {\bibinfo  {journal} {J. Phys. Chem. Lett.}\ }\textbf {\bibinfo {volume} {13}},\ \bibinfo {pages} {11503} (\bibinfo {year} {2022})}\BibitemShut {NoStop}%
\bibitem [{\citenamefont {Von~Neumann}(2018)}]{von2018}%
  \BibitemOpen
  \bibfield  {author} {\bibinfo {author} {\bibfnamefont {J.}~\bibnamefont {Von~Neumann}},\ }\href@noop {} {\emph {\bibinfo {title} {Mathematical foundations of quantum mechanics: New edition}}},\ Vol.~\bibinfo {volume} {53}\ (\bibinfo  {publisher} {Princeton university press},\ \bibinfo {year} {2018})\BibitemShut {NoStop}%
\bibitem [{\citenamefont {Misra}\ and\ \citenamefont {Sudarshan}(1977)}]{misra1977}%
  \BibitemOpen
  \bibfield  {author} {\bibinfo {author} {\bibfnamefont {B.}~\bibnamefont {Misra}}\ and\ \bibinfo {author} {\bibfnamefont {E.~G.}\ \bibnamefont {Sudarshan}},\ }\bibfield  {title} {\bibinfo {title} {The zeno's paradox in quantum theory},\ }\href@noop {} {\bibfield  {journal} {\bibinfo  {journal} {J. Math. Phys}\ }\textbf {\bibinfo {volume} {18}},\ \bibinfo {pages} {756} (\bibinfo {year} {1977})}\BibitemShut {NoStop}%
\bibitem [{\citenamefont {Facchi}\ and\ \citenamefont {Pascazio}(2008)}]{facchi2008}%
  \BibitemOpen
  \bibfield  {author} {\bibinfo {author} {\bibfnamefont {P.}~\bibnamefont {Facchi}}\ and\ \bibinfo {author} {\bibfnamefont {S.}~\bibnamefont {Pascazio}},\ }\bibfield  {title} {\bibinfo {title} {Quantum zeno dynamics: mathematical and physical aspects},\ }\href@noop {} {\bibfield  {journal} {\bibinfo  {journal} {J. Phys. A: Math. Theor.}\ }\textbf {\bibinfo {volume} {41}},\ \bibinfo {pages} {493001} (\bibinfo {year} {2008})}\BibitemShut {NoStop}%
\bibitem [{\citenamefont {Gu}\ and\ \citenamefont {Franco}(2017)}]{Bing2017}%
  \BibitemOpen
  \bibfield  {author} {\bibinfo {author} {\bibfnamefont {B.}~\bibnamefont {Gu}}\ and\ \bibinfo {author} {\bibfnamefont {I.}~\bibnamefont {Franco}},\ }\bibfield  {title} {\bibinfo {title} {Quantifying early time quantum decoherence dynamics through fluctuations},\ }\href {https://doi.org/10.1021/acs.jpclett.7b01817} {\bibfield  {journal} {\bibinfo  {journal} {J. Phys. Chem. Lett.}\ }\textbf {\bibinfo {volume} {8}},\ \bibinfo {pages} {4289} (\bibinfo {year} {2017})}\BibitemShut {NoStop}%
\bibitem [{\citenamefont {Kubo}(1962)}]{kubo1962}%
  \BibitemOpen
  \bibfield  {author} {\bibinfo {author} {\bibfnamefont {R.}~\bibnamefont {Kubo}},\ }\bibfield  {title} {\bibinfo {title} {A stochastic theory of line-shape and relaxation},\ }\href@noop {} {\bibfield  {journal} {\bibinfo  {journal} {Fluctuation, Relaxation and Resonance in Magnetic Systems}\ }\textbf {\bibinfo {volume} {23}} (\bibinfo {year} {1962})}\BibitemShut {NoStop}%
\bibitem [{\citenamefont {Yan}\ and\ \citenamefont {Zurek}(2022)}]{yan2022}%
  \BibitemOpen
  \bibfield  {author} {\bibinfo {author} {\bibfnamefont {B.}~\bibnamefont {Yan}}\ and\ \bibinfo {author} {\bibfnamefont {W.~H.}\ \bibnamefont {Zurek}},\ }\bibfield  {title} {\bibinfo {title} {Decoherence factor as a convolution: an interplay between a gaussian and an exponential coherence loss},\ }\href@noop {} {\bibfield  {journal} {\bibinfo  {journal} {New J. Phys.}\ }\textbf {\bibinfo {volume} {24}},\ \bibinfo {pages} {113029} (\bibinfo {year} {2022})}\BibitemShut {NoStop}%
\bibitem [{\citenamefont {Page}\ and\ \citenamefont {Tonks}(1981)}]{Page1981Separation}%
  \BibitemOpen
  \bibfield  {author} {\bibinfo {author} {\bibfnamefont {J.~B.}\ \bibnamefont {Page}}\ and\ \bibinfo {author} {\bibfnamefont {D.~L.}\ \bibnamefont {Tonks}},\ }\bibfield  {title} {\bibinfo {title} {On the separation of resonance raman scattering into orders in the time correlator theory},\ }\href {https://doi.org/10.1063/1.442006} {\bibfield  {journal} {\bibinfo  {journal} {J. Chem. Phys.}\ }\textbf {\bibinfo {volume} {75}},\ \bibinfo {pages} {5694} (\bibinfo {year} {1981})}\BibitemShut {NoStop}%
\bibitem [{\citenamefont {Shreve}\ and\ \citenamefont {Mathies}(1995)}]{shreve1995Thermal}%
  \BibitemOpen
  \bibfield  {author} {\bibinfo {author} {\bibfnamefont {A.~P.}\ \bibnamefont {Shreve}}\ and\ \bibinfo {author} {\bibfnamefont {R.~A.}\ \bibnamefont {Mathies}},\ }\bibfield  {title} {\bibinfo {title} {Thermal effects in resonance raman scattering: Analysis of the raman intensities of rhodopsin and of the time-resolved raman scattering of bacteriorhodopsin},\ }\href {https://doi.org/10.1021/j100019a012} {\bibfield  {journal} {\bibinfo  {journal} {J. Phys. Chem.}\ }\textbf {\bibinfo {volume} {99}},\ \bibinfo {pages} {7285} (\bibinfo {year} {1995})}\BibitemShut {NoStop}%
\bibitem [{\citenamefont {Tannor}\ and\ \citenamefont {Heller}(1982)}]{Tannor1982Poly}%
  \BibitemOpen
  \bibfield  {author} {\bibinfo {author} {\bibfnamefont {D.~J.}\ \bibnamefont {Tannor}}\ and\ \bibinfo {author} {\bibfnamefont {E.~J.}\ \bibnamefont {Heller}},\ }\bibfield  {title} {\bibinfo {title} {Polyatomic raman scattering for general harmonic potentials},\ }\href {https://doi.org/10.1063/1.443643} {\bibfield  {journal} {\bibinfo  {journal} {J. Chem. Phys.}\ }\textbf {\bibinfo {volume} {77}},\ \bibinfo {pages} {202} (\bibinfo {year} {1982})}\BibitemShut {NoStop}%
\bibitem [{\citenamefont {Myers}\ \emph {et~al.}(1982)\citenamefont {Myers}, \citenamefont {Mathies}, \citenamefont {Tannor},\ and\ \citenamefont {Heller}}]{Myers1982Excited}%
  \BibitemOpen
  \bibfield  {author} {\bibinfo {author} {\bibfnamefont {A.~B.}\ \bibnamefont {Myers}}, \bibinfo {author} {\bibfnamefont {R.~A.}\ \bibnamefont {Mathies}}, \bibinfo {author} {\bibfnamefont {D.~J.}\ \bibnamefont {Tannor}},\ and\ \bibinfo {author} {\bibfnamefont {E.~J.}\ \bibnamefont {Heller}},\ }\bibfield  {title} {\bibinfo {title} {Excited state geometry changes from preresonance raman intensities: isoprene and hexatriene},\ }\href@noop {} {\bibfield  {journal} {\bibinfo  {journal} {J. Chem. Phys.}\ }\textbf {\bibinfo {volume} {77}},\ \bibinfo {pages} {3857} (\bibinfo {year} {1982})}\BibitemShut {NoStop}%
\bibitem [{Note1()}]{Note1}%
  \BibitemOpen
  \bibinfo {note} {Classical noise processes have been extensively used to model line-broadening functions in spectroscopy. \cite {Mukamel_book,kubo1962,kubo1969,tokmakotime} In this model, the effect of the bath is incorporated by letting the electronic energy gap, $\omega _{01}$, fluctuate in time $\omega _{01} \to \omega _{01} + \delta \omega _{01}(t)$, where $\delta \omega _{01}(t)$ is a random function of time given by a Gaussian distribution. The resulting classical frequency correlation function is given by \begin {equation} \begin {aligned} {C_{\delta \omega _{01}}(t)}&{=\left \langle \delta \omega _{01}(t) \delta \omega _{01}(0) \right \rangle } \\ &{=\Delta ^{2} \exp \left [-t/\tau _{c} \right ].} \end {aligned} \end {equation} Here, $\Delta = \protect \sqrt {\left \langle \delta \omega _{01}^{2}\right \rangle }$ is the variance of the Gaussian distribution of frequencies and $\tau _{c}$ is the correlation time. The resulting, classical $g_{c}(t)$, line-broadening function is\cite
  {Mukamel_book,tokmakotime} \begin {equation} \begin {aligned} {g_{c}(t)} & {= - \DOTSI \intop \ilimits@ _{0}^{t}d\tau \protect \, (t-\tau ) \left \langle \delta \omega _{01}(\tau ) \delta \omega _{01}(0) \right \rangle } \\ &{= \Delta ^{2} \tau _{c}^{2} \left ( \exp (-t/\tau _{c})+t/\tau _{c}-1 \right ).} \end {aligned} \end {equation} It is important to note that the classical line-broadening function is purely real. Thus, it does not capture Stokes shifts and does not satisfy the fluctuation-dissipation theorem.\cite {Mukamel_book} In the limit where $t \ll \tau _{c}$ the classical correlation function becomes a constant, $C_{\delta \omega _{01}}(t)\approx \Delta ^{2}$, and the line-broadening function yields $g_{c}(t)=\Delta ^{2}t^{2}/2$. Because the correlation function is a constant, this is known as the inhomogeneous limit. Note that if $\Delta ^{2}=\DOTSI \intop \ilimits@ _{0}^{\infty } d\omega \protect \, J(\omega ) \coth {\left (\protect \frac {\omega } {2k_{B}T}\right )}$ we retrieve Eq.~\protect
  \eqref {eq:st-approx}. In turn, in the homogeneous limit, we have that $t \gg \tau _{c}$ and the classical line-broadening function becomes $g_{c}(t)=\Delta ^{2}\tau _{c}t$. Thus, we can mimic the real part of Eq.~\protect \eqref {eq:lt-limit} by setting $\Delta ^{2}\tau _{c} = \protect \mathcal {C}^{0}$, but we are completely missing the imaginary component. However, this does not mean that quantum and classical noise processes are equivalent but rather that in this limit classical noise processes can mimic quantum decoherence because the line broadening function is purely real, as previously noted in the literature.\cite {skinner1986pure,Bing2017,Vezvaee2024}.}\BibitemShut {Stop}%
\bibitem [{\citenamefont {Nian}\ \emph {et~al.}(2024)\citenamefont {Nian}, \citenamefont {Vinograd}, \citenamefont {Green}, \citenamefont {Chaffey}, \citenamefont {Massat}, \citenamefont {Singh}, \citenamefont {Zic}, \citenamefont {Fisher},\ and\ \citenamefont {Curro}}]{nian2024spin}%
  \BibitemOpen
  \bibfield  {author} {\bibinfo {author} {\bibfnamefont {Y.}~\bibnamefont {Nian}}, \bibinfo {author} {\bibfnamefont {I.}~\bibnamefont {Vinograd}}, \bibinfo {author} {\bibfnamefont {T.}~\bibnamefont {Green}}, \bibinfo {author} {\bibfnamefont {C.}~\bibnamefont {Chaffey}}, \bibinfo {author} {\bibfnamefont {P.}~\bibnamefont {Massat}}, \bibinfo {author} {\bibfnamefont {R.}~\bibnamefont {Singh}}, \bibinfo {author} {\bibfnamefont {M.}~\bibnamefont {Zic}}, \bibinfo {author} {\bibfnamefont {I.}~\bibnamefont {Fisher}},\ and\ \bibinfo {author} {\bibfnamefont {N.}~\bibnamefont {Curro}},\ }\bibfield  {title} {\bibinfo {title} {Spin echo, fidelity, and the quantum critical fan in tmvo 4},\ }\href@noop {} {\bibfield  {journal} {\bibinfo  {journal} {Phys. Rev. Lett.}\ }\textbf {\bibinfo {volume} {132}},\ \bibinfo {pages} {216502} (\bibinfo {year} {2024})}\BibitemShut {NoStop}%
\bibitem [{\citenamefont {Cucchietti}\ \emph {et~al.}(2005)\citenamefont {Cucchietti}, \citenamefont {Paz},\ and\ \citenamefont {Zurek}}]{cucchietti2005}%
  \BibitemOpen
  \bibfield  {author} {\bibinfo {author} {\bibfnamefont {F.}~\bibnamefont {Cucchietti}}, \bibinfo {author} {\bibfnamefont {J.~P.}\ \bibnamefont {Paz}},\ and\ \bibinfo {author} {\bibfnamefont {W.}~\bibnamefont {Zurek}},\ }\bibfield  {title} {\bibinfo {title} {Decoherence from spin environments},\ }\href@noop {} {\bibfield  {journal} {\bibinfo  {journal} {Phys. Rev. A}\ }\textbf {\bibinfo {volume} {72}},\ \bibinfo {pages} {052113} (\bibinfo {year} {2005})}\BibitemShut {NoStop}%
\bibitem [{\citenamefont {Zurek}\ \emph {et~al.}(2007)\citenamefont {Zurek}, \citenamefont {Cucchietti},\ and\ \citenamefont {Paz}}]{zurek2007echo}%
  \BibitemOpen
  \bibfield  {author} {\bibinfo {author} {\bibfnamefont {W.~H.}\ \bibnamefont {Zurek}}, \bibinfo {author} {\bibfnamefont {F.~M.}\ \bibnamefont {Cucchietti}},\ and\ \bibinfo {author} {\bibfnamefont {J.~P.}\ \bibnamefont {Paz}},\ }\bibfield  {title} {\bibinfo {title} {Gaussian decoherence and gaussian echo from spin environments},\ }\href@noop {} {\bibfield  {journal} {\bibinfo  {journal} {Acta Phys. Polym., B}\ }\textbf {\bibinfo {volume} {38}},\ \bibinfo {pages} {1685} (\bibinfo {year} {2007})}\BibitemShut {NoStop}%
\bibitem [{\citenamefont {Yarasi}\ \emph {et~al.}(2007)\citenamefont {Yarasi}, \citenamefont {Brost},\ and\ \citenamefont {Loppnow}}]{yarasi2007}%
  \BibitemOpen
  \bibfield  {author} {\bibinfo {author} {\bibfnamefont {S.}~\bibnamefont {Yarasi}}, \bibinfo {author} {\bibfnamefont {P.}~\bibnamefont {Brost}},\ and\ \bibinfo {author} {\bibfnamefont {G.~R.}\ \bibnamefont {Loppnow}},\ }\bibfield  {title} {\bibinfo {title} {Initial excited-state structural dynamics of thymine are coincident with the expected photochemical dynamics},\ }\href@noop {} {\bibfield  {journal} {\bibinfo  {journal} {J. Phys. Chem. A}\ }\textbf {\bibinfo {volume} {111}},\ \bibinfo {pages} {5130} (\bibinfo {year} {2007})}\BibitemShut {NoStop}%
\bibitem [{\citenamefont {Billinghurst}\ \emph {et~al.}(2012)\citenamefont {Billinghurst}, \citenamefont {Oladepo},\ and\ \citenamefont {Loppnow}}]{billinghurst2012}%
  \BibitemOpen
  \bibfield  {author} {\bibinfo {author} {\bibfnamefont {B.~E.}\ \bibnamefont {Billinghurst}}, \bibinfo {author} {\bibfnamefont {S.~A.}\ \bibnamefont {Oladepo}},\ and\ \bibinfo {author} {\bibfnamefont {G.~R.}\ \bibnamefont {Loppnow}},\ }\bibfield  {title} {\bibinfo {title} {{Initial excited-state structural dynamics of thymine derivatives}},\ }\href {https://doi.org/10.1021/jp301952v} {\bibfield  {journal} {\bibinfo  {journal} {J. Phys. Chem. B}\ }\textbf {\bibinfo {volume} {116}},\ \bibinfo {pages} {10496} (\bibinfo {year} {2012})}\BibitemShut {NoStop}%
\bibitem [{\citenamefont {Shu}\ and\ \citenamefont {Truhlar}(2023)}]{shu2023}%
  \BibitemOpen
  \bibfield  {author} {\bibinfo {author} {\bibfnamefont {Y.}~\bibnamefont {Shu}}\ and\ \bibinfo {author} {\bibfnamefont {D.~G.}\ \bibnamefont {Truhlar}},\ }\bibfield  {title} {\bibinfo {title} {Decoherence and its role in electronically nonadiabatic dynamics},\ }\href@noop {} {\bibfield  {journal} {\bibinfo  {journal} {J. Chem. Theory Comput.}\ }\textbf {\bibinfo {volume} {19}},\ \bibinfo {pages} {380} (\bibinfo {year} {2023})}\BibitemShut {NoStop}%
\bibitem [{\citenamefont {Izmaylov}\ and\ \citenamefont {Franco}(2017)}]{izmaylov2017entanglement}%
  \BibitemOpen
  \bibfield  {author} {\bibinfo {author} {\bibfnamefont {A.~F.}\ \bibnamefont {Izmaylov}}\ and\ \bibinfo {author} {\bibfnamefont {I.}~\bibnamefont {Franco}},\ }\bibfield  {title} {\bibinfo {title} {Entanglement in the born--oppenheimer approximation},\ }\href@noop {} {\bibfield  {journal} {\bibinfo  {journal} {J. Chem. Theory Comput}\ }\textbf {\bibinfo {volume} {13}},\ \bibinfo {pages} {20} (\bibinfo {year} {2017})}\BibitemShut {NoStop}%
\bibitem [{\citenamefont {Chang}\ \emph {et~al.}(2019)\citenamefont {Chang}, \citenamefont {Balciunas}, \citenamefont {Yin}, \citenamefont {Sapunar}, \citenamefont {Tenorio}, \citenamefont {Paul}, \citenamefont {Tsuru}, \citenamefont {Koch}, \citenamefont {Wolf}, \citenamefont {Coriani},\ and\ \citenamefont {W\"orner}}]{Marceau2019}%
  \BibitemOpen
  \bibfield  {author} {\bibinfo {author} {\bibfnamefont {Y.-P.}\ \bibnamefont {Chang}}, \bibinfo {author} {\bibfnamefont {T.}~\bibnamefont {Balciunas}}, \bibinfo {author} {\bibfnamefont {Z.}~\bibnamefont {Yin}}, \bibinfo {author} {\bibfnamefont {M.}~\bibnamefont {Sapunar}}, \bibinfo {author} {\bibfnamefont {B.~N.~C.}\ \bibnamefont {Tenorio}}, \bibinfo {author} {\bibfnamefont {A.~C.}\ \bibnamefont {Paul}}, \bibinfo {author} {\bibfnamefont {S.}~\bibnamefont {Tsuru}}, \bibinfo {author} {\bibfnamefont {H.}~\bibnamefont {Koch}}, \bibinfo {author} {\bibfnamefont {J.-P.}\ \bibnamefont {Wolf}}, \bibinfo {author} {\bibfnamefont {S.}~\bibnamefont {Coriani}},\ and\ \bibinfo {author} {\bibfnamefont {H.~J.}\ \bibnamefont {W\"orner}},\ }\bibfield  {title} {\bibinfo {title} {Non-born-oppenheimer electronic wave packet in molecular nitrogen at 14 ev probed by time-resolved photoelectron spectroscopy},\ }\href {https://doi.org/10.1103/PhysRevA.99.023426} {\bibfield  {journal} {\bibinfo  {journal} {Phys. Rev. A}\ }\textbf
  {\bibinfo {volume} {99}},\ \bibinfo {pages} {023426} (\bibinfo {year} {2019})}\BibitemShut {NoStop}%
\bibitem [{\citenamefont {Chang}\ \emph {et~al.}(2024)\citenamefont {Chang}, \citenamefont {Balciunas}, \citenamefont {Yin}, \citenamefont {Sapunar}, \citenamefont {Tenorio}, \citenamefont {Paul}, \citenamefont {Tsuru}, \citenamefont {Koch}, \citenamefont {Wolf}, \citenamefont {Coriani} \emph {et~al.}}]{Chang2024}%
  \BibitemOpen
  \bibfield  {author} {\bibinfo {author} {\bibfnamefont {Y.-P.}\ \bibnamefont {Chang}}, \bibinfo {author} {\bibfnamefont {T.}~\bibnamefont {Balciunas}}, \bibinfo {author} {\bibfnamefont {Z.}~\bibnamefont {Yin}}, \bibinfo {author} {\bibfnamefont {M.}~\bibnamefont {Sapunar}}, \bibinfo {author} {\bibfnamefont {B.~N.}\ \bibnamefont {Tenorio}}, \bibinfo {author} {\bibfnamefont {A.~C.}\ \bibnamefont {Paul}}, \bibinfo {author} {\bibfnamefont {S.}~\bibnamefont {Tsuru}}, \bibinfo {author} {\bibfnamefont {H.}~\bibnamefont {Koch}}, \bibinfo {author} {\bibfnamefont {J.-P.}\ \bibnamefont {Wolf}}, \bibinfo {author} {\bibfnamefont {S.}~\bibnamefont {Coriani}}, \emph {et~al.},\ }\bibfield  {title} {\bibinfo {title} {Electronic dynamics created at conical intersections and its dephasing in aqueous solution},\ }\href@noop {} {\bibfield  {journal} {\bibinfo  {journal} {Nat. Phys.}\ ,\ \bibinfo {pages} {1}} (\bibinfo {year} {2024})}\BibitemShut {NoStop}%
\bibitem [{\citenamefont {Hu}\ \emph {et~al.}(2020)\citenamefont {Hu}, \citenamefont {Gu},\ and\ \citenamefont {Franco}}]{Wen_2020}%
  \BibitemOpen
  \bibfield  {author} {\bibinfo {author} {\bibfnamefont {W.}~\bibnamefont {Hu}}, \bibinfo {author} {\bibfnamefont {B.}~\bibnamefont {Gu}},\ and\ \bibinfo {author} {\bibfnamefont {I.}~\bibnamefont {Franco}},\ }\bibfield  {title} {\bibinfo {title} {Toward the laser control of electronic decoherence},\ }\href {https://doi.org/10.1063/5.0002166} {\bibfield  {journal} {\bibinfo  {journal} {J. Chem. Phys.}\ }\textbf {\bibinfo {volume} {152}},\ \bibinfo {pages} {184305} (\bibinfo {year} {2020})}\BibitemShut {NoStop}%
\bibitem [{\citenamefont {Jasper}\ and\ \citenamefont {Truhlar}(2005)}]{jasper2005electronic}%
  \BibitemOpen
  \bibfield  {author} {\bibinfo {author} {\bibfnamefont {A.~W.}\ \bibnamefont {Jasper}}\ and\ \bibinfo {author} {\bibfnamefont {D.~G.}\ \bibnamefont {Truhlar}},\ }\bibfield  {title} {\bibinfo {title} {Electronic decoherence time for non-born-oppenheimer trajectories},\ }\href@noop {} {\bibfield  {journal} {\bibinfo  {journal} {J. Chem. Phys.}\ }\textbf {\bibinfo {volume} {123}} (\bibinfo {year} {2005})}\BibitemShut {NoStop}%
\bibitem [{\citenamefont {Jasper}\ \emph {et~al.}(2006)\citenamefont {Jasper}, \citenamefont {Nangia}, \citenamefont {Zhu},\ and\ \citenamefont {Truhlar}}]{jasper2006non}%
  \BibitemOpen
  \bibfield  {author} {\bibinfo {author} {\bibfnamefont {A.~W.}\ \bibnamefont {Jasper}}, \bibinfo {author} {\bibfnamefont {S.}~\bibnamefont {Nangia}}, \bibinfo {author} {\bibfnamefont {C.}~\bibnamefont {Zhu}},\ and\ \bibinfo {author} {\bibfnamefont {D.~G.}\ \bibnamefont {Truhlar}},\ }\bibfield  {title} {\bibinfo {title} {Non-born- oppenheimer molecular dynamics},\ }\href@noop {} {\bibfield  {journal} {\bibinfo  {journal} {Acc. Chem. Res.}\ }\textbf {\bibinfo {volume} {39}},\ \bibinfo {pages} {101} (\bibinfo {year} {2006})}\BibitemShut {NoStop}%
\end{thebibliography}%

\newpage
\clearpage % Force all floats to be placed

\begin{figure*}
    \centering
    \includegraphics[width=\textwidth]{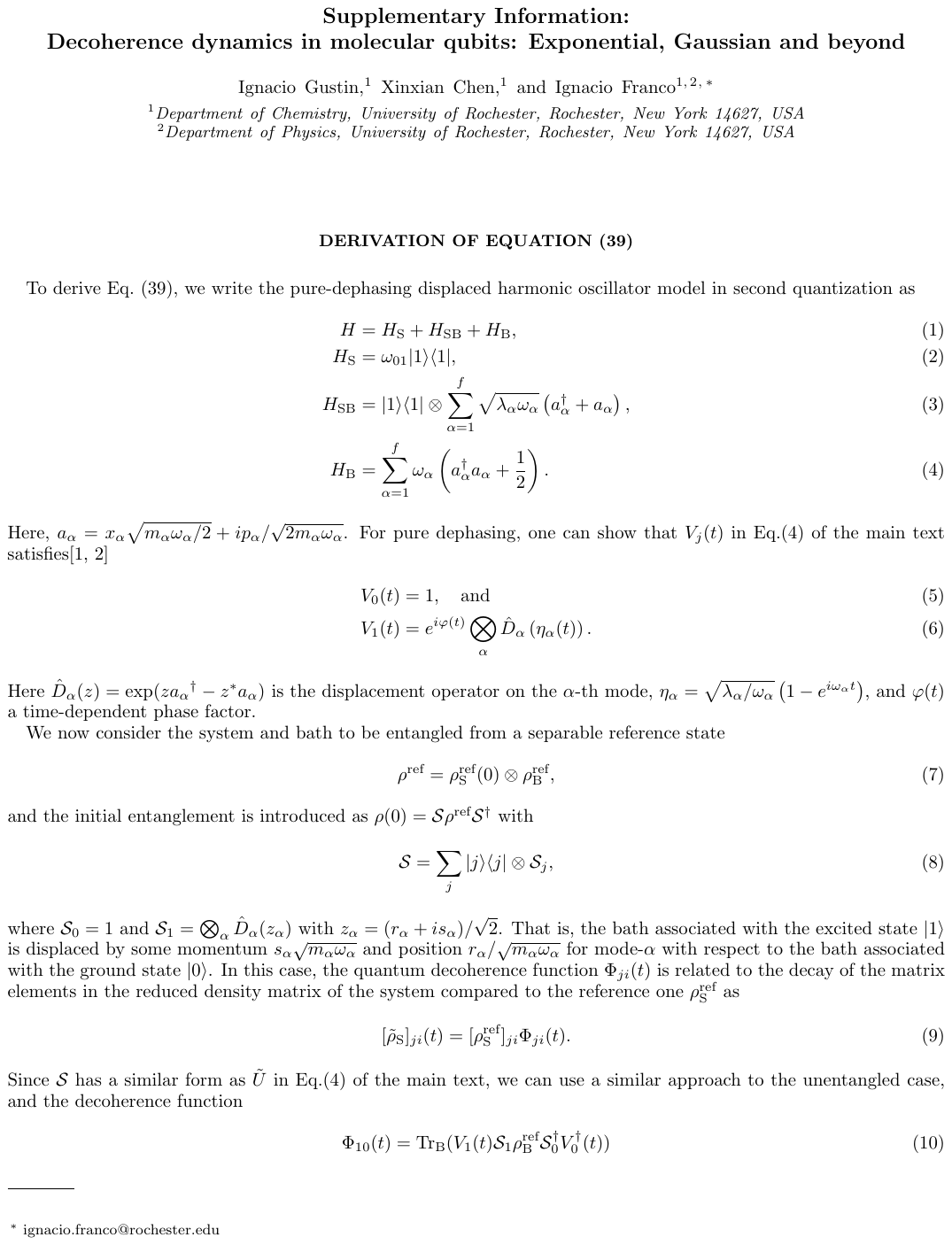} % <--- Key change here
\end{figure*}

\begin{figure*}
    \centering
    \includegraphics[width=\textwidth]{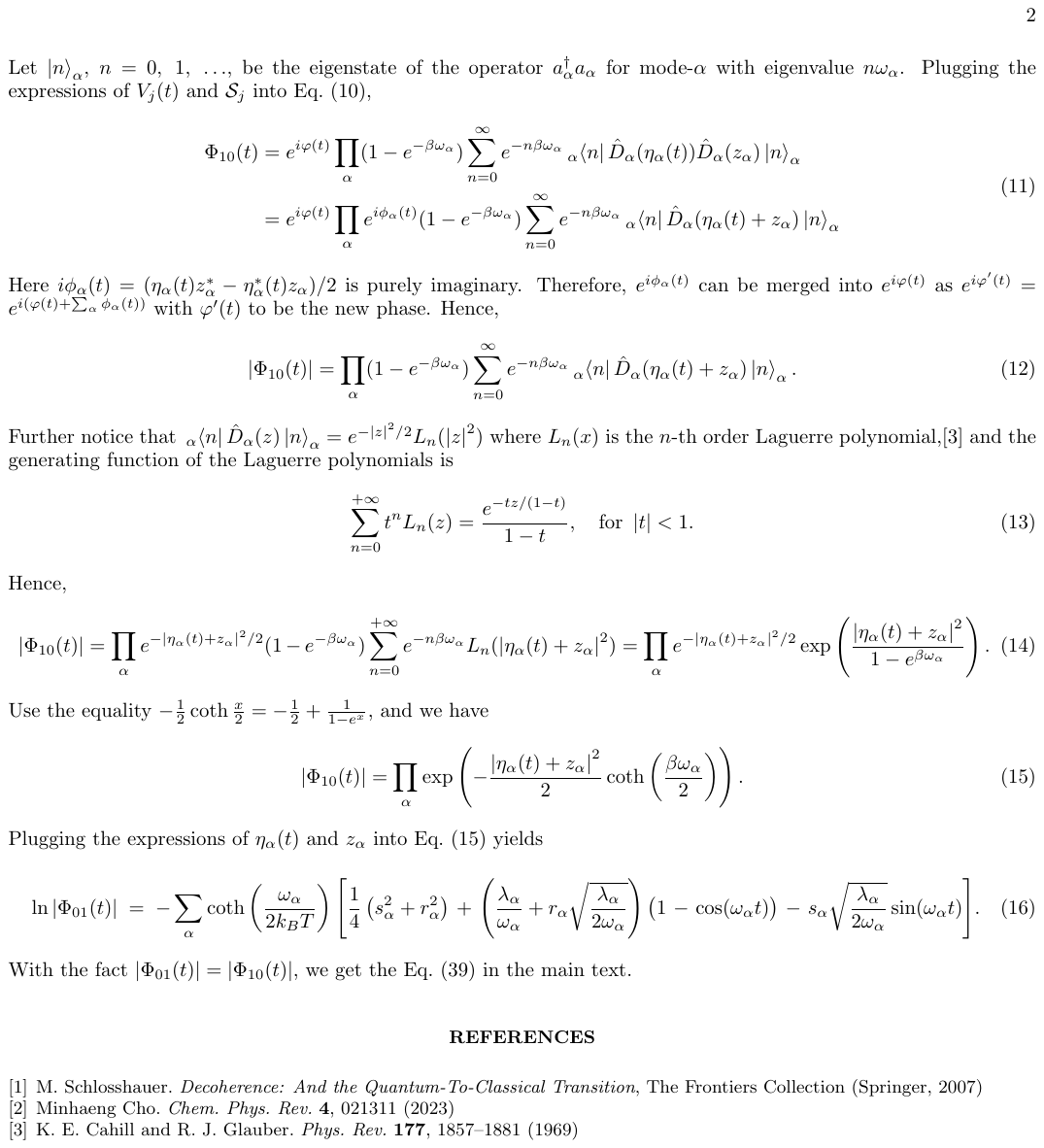} % <--- Key change here
\end{figure*}

\end{document}